\documentclass[sigconf]{acmart}

\usepackage{nicefrac}
\usepackage{siunitx}
\usepackage{array,framed}
\usepackage{booktabs}
\usepackage{
  color,
  float,
  epsfig,
  wrapfig,
  graphics,
  graphicx,
  subcaption
}
\usepackage{textcomp}
\usepackage{setspace}
\usepackage{latexsym,url}
\usepackage{enumerate}
\usepackage{enumitem}
\setenumerate[1]{label={\arabic*.}}
\usepackage{algorithm}
\usepackage{algpseudocode}
\usepackage{graphics}
\usepackage{xparse}
\usepackage{xspace}
\usepackage{multirow}
\usepackage{csvsimple}
\usepackage{balance}
\usepackage{circledsteps}
\pgfkeys{/csteps/inner color=white}
\pgfkeys{/csteps/fill color=black}

\usepackage{
  tikz,
  pgfplots,
  pgfplotstable
}
\usepackage{hyperref}
\usepackage[capitalise]{cleveref}

\usetikzlibrary{
  shapes.geometric,
  arrows,
  external,
  pgfplots.groupplots,
  matrix
}

\pgfplotsset{compat=1.9}

\usepackage{mathtools}

\DeclareMathAlphabet{\mathcal}{OMS}{cmsy}{m}{n}

\DeclareGraphicsExtensions{%
    .png,.PNG,%
    .pdf,.PDF,%
    .jpg,.mps,.jpeg,.jbig2,.jb2,.JPG,.JPEG,.JBIG2,.JB2}

\usepackage{xparse}
\usepackage[linewidth=0.5pt]{mdframed}
\newcommand{\ceil}[1]{{\lceil #1 \rceil}}

\newcommand{\tbeta}{\tilde{\beta}}
\newcommand{\tV}{\tilde{V}}

\newcommand{\tadd}{\tilde{add}}
\newcommand{\tmult}{\tilde{mult}}
\newcommand{\V}{\mathcal{V}}

\renewcommand{\P}{\mathcal{P}}
\renewcommand{\S}{\mathcal{S}}

\newcommand{\R}{\mathcal{R}}

\renewcommand{\varepsilon}{\mathcal{E}}

\newcommand{\accept}{{\tt 1}}
\newcommand{\reject}{{\tt 0}}
\newcommand{\pp}{\mathsf{pp}}

\newcommand{\MT}{\mathsf{MT}}

\newcommand{\vecr}{\mathbf{r}}
\newcommand{\vecx}{\mathbf{x}}
\newcommand{\vecy}{\mathbf{y}}
\newcommand{\vecf}{\mathbf{f}}
\newcommand{\vecq}{\mathbf{q}}

\newcommand{\vech}{\mathbf{h}}
\newcommand{\veci}{\mathbf{i}}

\newcommand{\vecb}{\mathbf{b}}
\newcommand{\vecz}{\mathbf{z}}
\newcommand{\vecg}{\mathbf{g}}
\newcommand{\vecu}{\mathbf{u}}
\newcommand{\vecv}{\mathbf{v}}
\newcommand{\vecw}{\mathbf{w}}
\newcommand{\vecone}{\mathbf{1}}

\newcommand{\PC}{\mathsf{PC}}

\newcommand{\com}{\mathsf{com}}

\newcommand{\KeyGen}{\mathsf{KeyGen}}
\newcommand{\Commit}{\mathsf{Commit}}

\newcommand{\Open}{\mathsf{Open}}
\newcommand{\Verify}{\mathsf{Verify}}

\newcommand{\rt}{\mathsf{rt}}

\newcommand{\bnm}{\begin{newmath}}
\newcommand{\enm}{\end{newmath}}

\newcommand{\bea}{\begin{eqnarray*}}%
\newcommand{\eea}{\end{eqnarray*}}%

\newcommand{\bne}{\begin{newequation}}
\newcommand{\ene}{\end{newequation}}

\newcommand{\bal}{\begin{newalign}}
\newcommand{\eal}{\end{newalign}}

\newenvironment{newalign}{\begin{align}%
\setlength{\abovedisplayskip}{4pt}%
\setlength{\belowdisplayskip}{4pt}%
\setlength{\abovedisplayshortskip}{6pt}%
\setlength{\belowdisplayshortskip}{6pt} }{\end{align}}

\newenvironment{newmath}{\begin{displaymath}%
\setlength{\abovedisplayskip}{4pt}%
\setlength{\belowdisplayskip}{4pt}%
\setlength{\abovedisplayshortskip}{6pt}%
\setlength{\belowdisplayshortskip}{6pt} }{\end{displaymath}}

\newenvironment{hproof}{
  \renewcommand{\proofname}{Proof (sketch)}\proof}{\endproof}
\newenvironment{newequation}{\begin{equation}%
\setlength{\abovedisplayskip}{4pt}%
\setlength{\belowdisplayskip}{4pt}%
\setlength{\abovedisplayshortskip}{6pt}%
\setlength{\belowdisplayshortskip}{6pt} }{\end{equation}}

\newcounter{ctr}

\newcounter{mytable}
\def\mytable{\begin{centering}\refstepcounter{mytable}}
\def\endmytable{\end{centering}}

\newcounter{myfig}
\def\myfig{\begin{centering}\refstepcounter{myfig}}
\def\endmyfig{\end{centering}}

\newlength{\saveparindent}
\newlength{\saveparskip}
\newcommand{\true}{\mathsf{true}}
\newcommand{\false}{\mathsf{false}}
\def\negl{\mathsf{negl}}
\newcommand{\updater}{updater~}

\usepackage{todonotes}
\newcounter{todocounter}

\newif\ifshowcomment
\showcommenttrue

\newcommand{\newcontent}[1]{{#1}}

\renewcommand{\eqref}[1]{\mbox{Equation~(\ref{#1})}}

\newcommand{\snark}{zk-SNARK}

\renewcommand{\vecw}{\mathbf{w}}

\def \part {part}

\renewcommand{\paragraph}[1]{\vspace{5pt}\noindent\textbf{#1}\;}

\def \blackslug{\hbox{\hskip 1pt \vrule width 4pt height 8pt
    depth 1.5pt \hskip 1pt}}
\def \qed{\quad\blackslug\lower 8.5pt\null\par}

\def \binary {\{0, 1\}}

\newcounter{mynote}[section]

\newcommand\ignore[1]{}

\newcounter{rcnote}[section]

\newcounter{mrnote}[section]

\newcounter{fknote}[section]

\newcounter{anote}[section]

\DeclareMathSymbol{\mlq}{\mathord}{operators}{``}
\DeclareMathSymbol{\mrq}{\mathord}{operators}{`'}

\newcommand{\rhf}[2]{R_{f, \gamma}}

\def\H{\ensuremath{\mathsf{H}}\xspace}

\def\F{\ensuremath{\mathsf{F}}\xspace}

\DeclareDocumentCommand{\edist}{o o}{
  \ensuremath{
    \IfNoValueTF{#1}{{d}}{{\sf d}(#1,#2)}
  }
}

\newcommand{\olrk}[1]{\ifx\nursymbol#1\else\!\!\mskip4.5mu plus 0.5mu\left(\mskip0.5mu plus0.5mu #1\mskip1.5mu plus0.5mu \right)\fi}

\NewDocumentCommand{\indseq}{ O{1} O{r} }{{#1}\ldots {#2}}

\newcommand{\circuit}{\mathsf{C}}

\newcommand{\block}{\mathsf{blk}}
\newcommand{\trx}{\mathsf{trx}}
\newcommand{\blocksize}{t}
\newcommand{\blklog}{\mathsf{LOG}}

\newcommand{\blockheader}{\mathsf{blkH}}
\newcommand{\blockpointer}{\mathsf{ptr}}
\newcommand{\lightclientcircuit}{\mathsf{LightCC}}

\newcommand{\numProver}{N}

\newcommand{\longestchainfinder}{\mathit{Longest}}

\setcopyright{none}
\renewcommand\footnotetextcopyrightpermission[1]{}

\newenvironment{ourprotocol}[1][htb]{%
    \refstepcounter{protocol}
    \floatname{algorithm}{Protocol}%
   \begin{algorithm}[#1]%
  }{\end{algorithm}}
\usepackage[subtle,tracking=normal, wordspacing=normal,mathdisplays=tight]{savetrees}

\newcommand{\systemname}{\textsf{zkBridge}\xspace}
\newcommand{\devirgo}{\textsf{deVirgo}\xspace}

\begin{document}
\title{\systemname: Trustless Cross-chain Bridges Made Practical}
\font\myfont=cmr12 at 9pt
\author{\myfont Tiancheng Xie$^1$, Jiaheng Zhang$^1$, Zerui Cheng$^2$, Fan Zhang$^3$, Yupeng Zhang$^4$, Yongzheng Jia$^7$, Dan Boneh$^5$, Dawn Song$^{1,6}$}

\begin{abstract}
  Blockchains have seen growing traction with cryptocurrencies reaching a market cap of over 1 trillion dollars, major institution investors taking interests, and global impacts on  governments, businesses, and individuals. 
Also growing significantly is the heterogeneity of the ecosystem where a variety of blockchains co-exist. 
Cross-chain bridge is a necessary building block in this multi-chain ecosystem.
Existing solutions, however, either suffer from performance issues or rely on trust assumptions of committees that significantly lower the security.
Recurring attacks against bridges have cost users more than 1.5 billion USD.
In this paper, we introduce $\systemname$, an efficient cross-chain bridge that guarantees strong security without external trust assumptions.
With succinct proofs,
$\systemname$ not only guarantees correctness, but also significantly reduces on-chain verification cost.
We propose novel succinct proof protocols that are orders-of-magnitude faster than existing solutions for workload in \systemname.
With a modular design, \systemname enables a broad spectrum of use cases and capabilities, including message passing, token transferring, and other computational logic operating on state changes from different chains.
To demonstrate the practicality of \systemname, we implemented a prototype bridge from Cosmos to Ethereum, a particularly challenging direction that involves large proof circuits that existing systems cannot efficiently handle. Our evaluation shows that $\systemname$ achieves practical performance: proof generation takes less than $20$ seconds, while verifying proofs on-chain costs less than 230K gas. For completeness, we also implemented and evaluated the direction from Ethereum to other EVM-compatible chains (such as BSC) which involves smaller circuits and incurs much less overhead.

\end{abstract}

\maketitle

\footnotetext[1]{UC Berkeley}
\footnotetext[2]{Tsinghua University}
\footnotetext[3]{Yale University}
\footnotetext[4]{Texas A\&M University}
\footnotetext[5]{Stanford University}
\footnotetext[6]{Oasis Labs}
\footnotetext[7]{Overeality Labs}

\pagestyle{plain}

\section{Introduction}
\label{sec:intro}

Since the debut of Bitcoin, blockchains have evolved to an expansive ecosystem of various applications and communities.
Cryptocurrencies like Bitcoin and Ethereum are gaining rapid traction with the market cap reaching over a trillion USD~\cite{coinmarketcap} and institutional investors~\cite{hamlin2022, wintermeyer2021} taking interests.
Decentralized Finance (DeFi) demonstrates that blockchains can enable finance instruments that are otherwise impossible (e.g., flash loans~\cite{qin2021attacking}).
More recently, digital artists~\cite{BeeplesoldanNFTfor69millionTheVerge-2022-04-24} and content creators~\cite{YouTubeincludesNFTsinnewcreatortools-2022-04-24} resort to blockchains for transparent and accountable circulation of their works.

Also growing significantly is the heterogeneity of the ecosystem.
A wide range of blockchains have been proposed and deployed, ranging from ones leveraging computation (e.g., in Proof-of-Work~\cite{nakamoto2008bitcoin}), to economic incentives (e.g., in Proof-of-Stake~\cite{algorand, coa, ouroboros, ouroborosbrao, snowwhite}), and various other resources such as storage~\cite{ren2016pos, filecoin, dziembowski2015pos, ateniese2014pos}, and even time~\cite{intelpoes}.
While it is rather unclear that one blockchain dominates others in all aspects,
these protocols employ different techniques and achieve different security guarantees and performance.
It has thus been envisioned that (e.g., in~\cite{Amultichainapproach,Multichainfuturelikely,buterin-multichain}) the ecosystem will grow to a {\em multi-chain} future where various protocols co-exist, and developers and users can choose the best blockchain based on their preferences, the cost, and the offered amenities.

\newcommand{\chainone}{\ensuremath{\mathcal{C}_1}}
\newcommand{\chaintwo}{\ensuremath{\mathcal{C}_2}}
\newcommand{\tokenone}{\ensuremath{\textsf{Token}_1}}
\newcommand{\tokentwo}{\ensuremath{\textsf{Token}_2}}
\newcommand{\contractone}{\ensuremath{\mathcal{SC}_1}}
\newcommand{\contracttwo}{\ensuremath{\mathcal{SC}_2}}
\newcommand{\vdep}{\texttt{\$v}}
\newcommand{\user}{\mathcal{U}}

A central challenge in the multi-chain universe is how to enable secure {\em cross-chain bridges} through which applications on different blockchains can communicate.
An ecosystem with efficient and inexpensive bridges will enable assets held on one chain to
effortlessly participate in marketplaces hosted on other chains.
In effect, an efficient system of bridges will do for blockchains what the Internet did for siloed communication networks. 

The core functionality of a bridge between blockchains $\chainone$ and $\chaintwo$ is to prove to applications on $\chaintwo$ that a certain event
took place on $\chainone$, and vice versa.
We use a generic notion of a bridge, namely one that can perform multiple functions: message passing, asset transfers, etc.  
In our modular design, the bridge itself neither involves nor is restricted to any application-specific logic.

\paragraph{The problem.}
While cross-chain bridges have been built in practice~\cite{rainbow, poly_network, layerzero, axelar}, existing solutions either suffer from poor performance, or rely on central parties.

The operation of the bridge depends on the consensus protocols of both chains.
If $\chainone$ runs Proof-of-Work, a natural idea is to use a light client protocol (e.g., SPV~\cite{nakamoto2008bitcoin}).
Specifically, a smart contract on $\chaintwo$, denoted by $\contracttwo$, will keep track of block headers of $\chainone$, based on which transaction inclusion (and other events) can be verified with Merkle proofs.
This approach, however, incurs a significant computation and storage overhead, since $\contracttwo$ needs to verify all block headers and keep a long and ever-growing list of them.
For non-PoW chains, the verification can be even more expensive.
For example, for a bridge between a Proof-of-Stake chain (like Cosmos) and Ethereum, verifying a single block header on Ethereum would cost about 64 million gas~\cite{nearbridge} (about \$6300 at time of writing),  which is prohibitively high.

Currently, as an efficient alternative, many bridge protocols (PolyNetwork, Wormhole, Ronin, etc.) resort to a committee-based approach: a committee of validators are entrusted to sign off on state transfers. 
In these systems, the security boils down to, e.g., the honest majority assumption.
This is problematic for two reasons.
First, the extra trust assumption in the committee means the bridged asset is not as secure as native ones, complicating the security analysis of downstream applications.
Second, relying on a small committee can lead to single point failures.
Indeed, in a recent exploit of the Ronin bridge~\cite{RoninAttackCoindesk}, the attackers were able to obtain five of the nine validator keys, through which they stole 624 million USD, making it the largest attack in the history of DeFi by Apr 2022\footnote{see the ranking at \url{https://rekt.news/leaderboard}}. Even the second and third largest attacks are also against bridges (\$611m was stolen from PolyNetwork~\cite{polynetworkattack} and \$326m was stolen from Wormhole~\cite{wormholeattack}), and key compromise was suspected in the PolyNetwork attack.

\paragraph{Our approach.} We present \systemname to enable an efficient cross-chain bridge without trusting a centralized committee. The main idea is to leverage \snark{}, which are succinct non-interactive proofs (arguments) of knowledge
\cite{hyrax, libra, virgoplus, libstark, libsnark, ligero, aurora, fractal, marlin,vsql, vram, bulletproofs, gabizon2019plonk, spartan}. 
A \snark{} enables a prover
to efficiently convince $\contracttwo$ that a certain state transition took place on $\chainone$. 
To do so, $\contracttwo$ will keep track of a digest $D$ of the latest tip of $\chainone$.
To sync $\contracttwo$ with new blocks in $\chainone$, 
anyone can generate and submit a \snark{} that proves to $\contracttwo$ that the tip of $\chainone$ has advanced from $D$ to $D'$.

This design offers three benefits.
First, the soundness property of a \snark{} ensures the security of the bridge.
Thus, we do not need additional security requirements beyond the security of the underlying blockchains.
In particular \systemname does not rely on a committee for security.
Second, with a purpose-built \snark{}, $\chaintwo$ can verify a state transition of $\chainone$ 
far more efficiently than encoding the consensus logic of $\chainone$ in $\contracttwo$.
In this way, as an example for  \systemname from Cosmos to Ethereum, we reduce the proof verification cost from $\sim80M$ gas to less than $230K$ gas on $\chaintwo$. 
The storage overhead of the bridge is reduced to constant. 
Third, by separating the bridge from application-specific logic, \systemname makes it easy to enable additional applications on top of the bridge.

\paragraph{Technical challenges.}
To prove correctness of a given computation outcome using a \snark{}, one first needs to express the computation as an arithmetic circuit.
While \snark{} verification is fast (logarithmic in the size of the circuit or even constant), proof generation time is at least linear, and in practice can be prohibitively expensive.
Moreover, components used by real-world blockchains are not easily expressed as an arithmetic circuit. 
For example, the widely used EdDSA digital signature scheme is very efficient to verify on a CPU, but is expensive to express as an arithmetic circuit, requiring more than 2 million gates~\cite{circom-ed25519}.
In a cross-chain bridge, each state transition could require the verification of hundreds of signatures depending on the chains, 
making it prohibitively expensive to generate the required \snark{} proof.
In order to make \systemname practical, we must reduce proof generation time.

To this end, we propose two novel ideas.
First, we observe that the circuits used by cross-chain bridges are {\em data-parallel}, in that they contain multiple identical copies of a smaller sub-circuit.
Specifically, the circuit for verifying $N$ digital signatures contains $N$ copies of the signature verification sub-circuit.
To leverage the data-parallelism, we propose \devirgo, a novel distributed zero-knowledge proofs protocol based on Virgo~\cite{virgo}. 
\devirgo enjoys perfect linear scalability, namely, the proof generation time can be reduced by a factor of $M$ if the generation is distributed over $M$ machines. 
The protocol is of independent interest and might be useful in other scenarios. Other proof systems can be similarly parallelized~\cite{wu2018dizk}.

While \devirgo significantly reduces the proof generation time, verifying \devirgo proofs on chain, 
especially for the billion-gate circuits in \systemname, can be expensive for smart contracts where computational resources are extremely limited.
To compress the proof size and the verification cost, 
we recursively prove the correctness of a (potentially large) $\devirgo$ proof using a classic \snark{} due to Groth~\cite{Groth2016}, hereafter denoted Groth16. 
The Groth16 prover outputs constant-size proofs that are fast to verify by a smart contract on an EVM blockchain.
We stress that one cannot use Groth16 to generate the entire \systemname proof because the circuits needed in \systemname are too large for a Groth16 prover. 
Instead, our approach of compressing a \devirgo proof using Groth16 gives the best of both worlds:
a fast \devirgo parallel prover for the bulk of the proof, where the resulting proof is compressed into a succinct Groth16 proof that is fast to verify. 
We elaborate on this technique in~\cref{sec:composition}.
This approach to compressing long proofs is also being adopted in commercial \snark{} systems such as~\cite{PolygonMiden, PolygonHermez, RiscZero}. 

\paragraph{Implementation and evaluation.} To demonstrate the practicality of \systemname, we implement an end-to-end prototype of $\systemname$ from Cosmos to Ethereum, given it is among the most challenging directions as it involves large circuits for correctness proofs. Our implementation includes the protocols of \devirgo and recursive proof with Groth16, and the transaction relay application. The experiments show that our system achieves practical performance.  \devirgo can generate a block header relay proof within 20s, which is more than 100x faster than the original Virgo system with a single machine. Additionally, the on-chain verification cost decreases from $\sim$80M gas (direct signature verification) to less than 230K gas, due to the recursive proofs.
In addition, as a prototype example, we also implement  zkBridge from Ethereum to other EVM-compatible chains such as BSC, which involves smaller circuits for proof generation and incurs much less overhead.

\subsection{Our contribution}
In this paper, we make the following contributions:

\begin{itemize}[leftmargin=*]
    \item In this paper, we propose \systemname, a trustless, efficient, and secure cross-chain bridge whose security relies on succinct proofs (cryptographic assumptions) rather than a committee (external trust assumptions). Compared with existing cross-chain bridge projects in the wild, \systemname is the first solution that achieves the following properties at the same time.
    \begin{itemize}
    
        \item \textbf{Trustless and Secure:} The correctness of block headers on remote blockchains is proven by zk-SNARKs, and thus no external trust assumptions are introduced. Indeed, as long as the connected blockchains and the underlying light-client protocols are secure, and there exists at least one honest node in the block header relay network, \systemname is secure.

        \item \textbf{Permissionless and Decentralized:} Any node can freely join the network to relay block headers, generate proofs, and claim the rewards. Due to the elimination of the commonly-used central or Proof-of-Stake style committee for block header validation, \systemname also enjoys better decentralization.
        
        \item \textbf{Extensible:} Smart contracts using \systemname enjoy maximum flexibility because they can invoke the updater contract to retrieve verified block headers, and then perform their application-specific verification and functionality (e.g., verifying transaction inclusion through auxiliary Merkle proofs). By separating the bridge from application-specific logic, \systemname makes it easy to develop  applications on top of the bridge.
        
         \item \textbf{Universal:}  The block header relay network and the underlying proof scheme in \systemname is universal as long as the blockchain supports a light client protocol as in Definition \ref{def:lightclient}.
         
         \item \textbf{Efficient:} With our highly optimized recursive proof scheme, block headers can be relayed within a short time (usually tens of seconds for proof generation), and the relayed information can be quickly finalized as soon as the proof is verified, thus supporting fast and flexible bridging of information.

    \end{itemize}
    
    In summary, zkBridge is a huge leap towards building a secure, trustless foundation for blockchain interoperability.
    
    \item We propose a novel 2-layer recursive proof system, which is of independent interest, as the underlying zk-SNARK protocol to achieve both reasonable proof generation time and on-chain verification cost. Through the coordination of \devirgo and Groth16, we achieve a desirable balance between efficiency and cost.
    \begin{itemize}
        \item For the first layer, aiming at prompt proof generation, we introduce \devirgo, a distributed version of Virgo proof system. \devirgo combines distributed sumcheck and distributed polynomial commitment to achieve optimal parallelism, through which the proof generation phase is much more accelerated by running on distributed machines. \devirgo is more than 100x faster than Virgo for the workload in \systemname. 
        \item  For the second layer, aiming at acceptable on-chain verification cost, we use Groth16 to recursively prove that the previously generated proof by \devirgo indeed proves the validity of the corresponding remote block headers. Through the second layer, the verification gas cost is reduced from an estimated $\sim 80M$ to less than $230K$, making on-chain verification practical. 
    \end{itemize}

    \item We implement an end-to-end prototype of \systemname and evaluate its performance in two scenarios: from Cosmos to Ethereum (which is the main focus since it involves large proof circuits that existing systems cannot efficiently handle), and from Ethereum to other EVM-compatible chains (which in comparison involves much smaller circuits). The experiment results show that \systemname achieves practical performance and is the first practical cross-chain bridge that achieves cryptographic assurance of correctness.
\end{itemize}

\renewcommand{\F}{\mathbb{F}}
\renewcommand{\H}{\mathbb{H}}
\renewcommand{\L}{\mathbb{L}}
\section{Background}
\label{sec:background}

In this section we cover the preliminaries, essential background on blockchains, and zero-knowledge proofs.

\subsection{Notations}
Let $\F$ be a finite field and $\lambda$ be a security parameter. We use $f(), h()$ for polynomials, $x, y$ for single variables, bold letters $\vecx, \vecy$ for vectors of variables.  Both $\vecx[i]$ and $x_i$ denote the $i$-th element in $\vecx$. For
$\vecx$, we use notation $\vecx[i:k]$ to denote slices of vector $\vecx$, namely $\vecx[i:k]=(x_i, x_{i+1}, \cdots, x_k)$. We use $\veci$ to denote the vector of the binary representation of some integer $i$.

\paragraph{Merkle Tree. } 
Merkle tree~\cite{merkletree} is a data structure widely used to build commitments to vectors because of its simplicity and efficiency. The prover time is linear in the size of the vector while the verifier time and proof size are logarithmic in the size of the vector. Given a vector of $\vecx = (x_0, \cdots, x_{N-1})$, it consists of three algorithms:
\begin{itemize}
    \item $\rt \leftarrow \MT.\Commit(\vecx)$
    \item $(\vecx[i], \pi_i) \leftarrow \MT.\Open( \vecx, i)$ \item $\{1, 0\} \leftarrow \MT.\Verify(\pi_i, \vecx[i], \rt)$.
\end{itemize}

\subsection{Blockchains}

A blockchain is a distributed protocol where a group of nodes collectively maintains a {\em ledger} which consists of an ordered list of {\em blocks}. 
A block $\block$ is a data-structure that stores a header $\blockheader$ and a list of transactions, denoted by $\block = \{\blockheader; \trx_1, \dots, \trx_{\blocksize}\}$.
A block header contains metadata about the block, including a pointer to the previous block, a compact representation of the transactions (typically a Merkle tree root), validity proofs such as solutions to cryptopuzzles in Proof-of-Work systems or validator signatures in Proof-of-Stake ones.

\paragraph{Security of blockchains.} The security of blockchains has been studied extensively.
Suppose the ledger in party $i$'s local view is $\blklog_i^r = [ \block_1, \block_2, \dots, \block_r]$ where $r$ is the {\em height}. For any $2 \le k \le r$ and the $k$-th block $\block_k$ , $\block_k.\blockpointer = \blockheader_{k-1}$, so every single block is linked to the previous one.
For the purpose of this paper, we care about two (informal) properties:

\begin{enumerate}
    \item {\bf Consistency}: For any honest nodes $i$ and $j$, and for any rounds of $r_0$ and $r_1$, it must be satisfied that either $\blklog_i^{r_0}$ is a prefix of  $\blklog_j^{r_1}$ or vice versa.
    \item {\bf Liveness}: If an honest node receives some transaction $\trx$ at some round $r$, then $\trx$ will be included into the blockchain of all honest nodes eventually.
\end{enumerate}

\paragraph{Smart contracts and gas.} 
In addition to reaching consensus over the content of the ledger, many blockchains support expressive user-defined programs called {\em smart contracts}, which are stateful programs with state persisted on a blockchain. 
Without loss of generality, smart contract states can be viewed a key-value store (and often implemented as such.)
Users send transactions to interact with a smart contract, and potentially alter its state.

A key limitation of existing smart contract platforms is that computation and storage are scarce resources and can be considerably expensive. 
Typically smart contract platforms such as Ethereum charge a fee (sometimes called gas) for every step of computation. 
For instance, 
EdDSA signatures are  extremely cheap to verify (a performant CPU can verify 71000 of them in a second~\cite{bernstein2012high}), but 
verifying a single EdDSA signature on Ethereum costs about 500K gas, which is about \$49 at the time of writing.
Storage is also expensive on Ethereum.
Storing 1KB of data costs about 0.032 ETH, which can be converted to approximately \$90 at the time of writing.
This limitation is not unique to Ethereum but rather a reflection of the low capacity of permissionless blockchains in general.
Therefore reducing on-chain computation and storage overhead is one of the key goals.

\subsection{Light client protocol} 
In a blockchain network, there are full nodes as well as light ones. 
Full nodes store the entire history of the blockchain and verify all transactions in addition to verifying  block headers. 
Light clients, on the other hand, only store the headers, and therefore can only verify a subset of correctness properties.

The workings of light clients depend on the underlying consensus protocol. 
The original Bitcoin paper contains a light client protocol (SPV~\cite{nakamoto2008bitcoin}) that uses Merkle proofs to enable a light client who only stores recent headers to verify transaction inclusion. A number of improvements have been proposed ever since. 
For instance, in Proof-of-Stake, typically a light client needs to verify account balances in the whole
blockchain history (or up to a snapshot), and considers the risk of long range attacks.
For BFT-based consensus, a light client needs to verify validator signatures and keeps track of validator rotation.
We refer readers to~\cite{chatzigiannis2021sok} for a survey.

\newcommand{\lcstate}{\textsf{LCS}}

To abstract consensus-specific details away, we use $$\lightclientcircuit(\lcstate_{r-1}, \blockheader_{r-1}, \blockheader_{r})\to \{\true,\false\}$$ to denote the block validation rule of a light client: given a new block header $\blockheader_r$, $\lightclientcircuit$ determines if the header represents a valid next block after $\blockheader_{r-1}$ given its current state $\lcstate_{r-1}$.
\newcontent{
We define the required properties of a light client protocol as follows:
\begin{definition}[Light client protocol]  \label{def:lightclient}
A light client protocol enables a node to synchronize the block headers of the state of the blockchain. Suppose all block headers in party $i$'s local view is $\blklog H_i^r = [\blockheader_{1}, \blockheader_{2}, ..., \blockheader_{r}]$, the light client protocol satisfies following properties:
\begin{enumerate}
    \item \textbf{Succinctness}: For each state update, the light client protocol only takes $O(1)$ time to synchronize the state.
    \item \textbf{Liveness}: If an honest full node receives some transaction $\trx$ at some round $r$, then $\trx$ must be included into the blockchain eventually. A light client protocol will eventually include a block header $\blockheader_{i}$ such that the corresponding block includes the transaction $\trx$.
    \item \textbf{Consistency}: 
    For any honest nodes $i$ and $j$, and for any rounds of $r_0$ and $r_1$, it must be satisfied that either $\blklog H_i^{r_0}$ is a prefix of $\blklog H_j^{r_1}$ or vice versa.
\end{enumerate}

\end{definition}
}
\subsection{Zero-knowledge proofs}

An argument system for an NP relationship $\mathcal{R}$ is a protocol between a computationally-bounded prover $\P$ and a verifier $\V$. At the end of the protocol, $\V$ is convinced by $\P$ that there exists a witness $\vecw$ such that $(\vecx; \vecw) \in \mathcal{R}$ for some input $\vecx$. We use $\mathcal{G}$ to represent the generation phase of the public parameters $\pp$. Formally, consider the definition below, where we assume $\mathcal{R}$ is known to $\P$ and $\V$.

\begin{definition}\label{def::zkp}
	
	Let $\lambda$ be a security parameter and $\mathcal{R}$ be an NP relation. A tuple of algorithm $(\mathcal{G}, \mathcal{P}, \mathcal{V})$ is a zero-knowledge argument of knowledge for $\mathcal{R}$ if the following holds.
	
	\begin{itemize}
		
		\item \textbf{Completeness}. For every $\pp$ output by $\mathcal{G}(1^\lambda)$, $(\vecx; \vecw) \in \mathcal{R}$ and $\pi \leftarrow \P(\vecx, \vecw, \pp)$, 
		$$\Pr[\V(\vecx, \pi, \pp) = \accept] = 1$$

		\item \textbf{Knowledge Soundness}. For any PPT prover $\P^*$, there exists a PPT extractor $\varepsilon$ such that for any auxiliary string $\vecz$, $\pp \leftarrow \mathcal{G}(1^\lambda)$, $\pi^* \leftarrow \P^*(\vecx, \vecz, \pp)$, $w \leftarrow \varepsilon^{\P^*(\cdot)}(\vecx, \vecz,  \pp)$, and
		\begin{align*}
		&\Pr[(\vecx; \vecw) \notin \mathcal{R} \wedge \V(\vecx, \pi^*, \pp) = \accept] \leq \negl(\lambda),
		\end{align*}
		where $\varepsilon^{\P^*(\cdot)}$ represents that $\varepsilon$ can rewind $\P^*$, 
		\item \textbf{Zero knowledge}. There exists a PPT simulator $\S$ such that for any PPT algorithm $\V^*$, $(\vecx;\vecw)\in \mathcal{R}$, $\pp$ output by $\mathcal{G}(1^\lambda)$, it holds that %
		\[
		\mathsf{View}(\V^*(\pp, \vecx)) \approx \S^{\V^*}(\vecx),
		\]
		where $\mathsf{View}(\V^*(\pp, \vecx))$ denotes the view that the verifier sees during the execution of the interactive process with $\P$, $\S^{\V^*}(\vecx)$ denotes the view generated by $\S$ given input $\vecx$ and transcript of $\V^*$, and $\approx$ denotes two perfectly indistinguishable distributions.
		
	\end{itemize}
	We say that $(\mathcal{G},\P,\V)$ is a \textbf{succinct} argument system\footnote{In our construction, we only need a succinct non-interactive arguments of knowledge (SNARK) satisfying the first two properties and the succinctness for validity. The zero knowledge property could be used to further achieve privacy.} if the total communication (proof size) between $\P$ and $\V$,
	as well as $\V$'s running time, are $\mathsf{poly}(\lambda,|\vecx|,\log|\mathcal{R}|)$, where $|\mathcal{R}|$ is the size of the circuit that computes $\mathcal{R}$ as a function of $\lambda$.
\end{definition}
\section{\systemname Protocol}

\label{sec:overview}

At a high level, a smart contract is a stateful program with states persisted on a blockchain.
A bridge like \systemname is a service that enables smart contracts on different blockchains to transfer states from one chain to another in a secure and verifiable fashion. 

Below we first explain the design of \systemname and its workflow through an example, then we specify the protocol in more detail. For ease of exposition, we focus on one direction of the bridge, but the operation of the opposite direction is symmetric. 

\begin{figure*}[ht]
    \centering
    \includegraphics[width=0.73\textwidth]{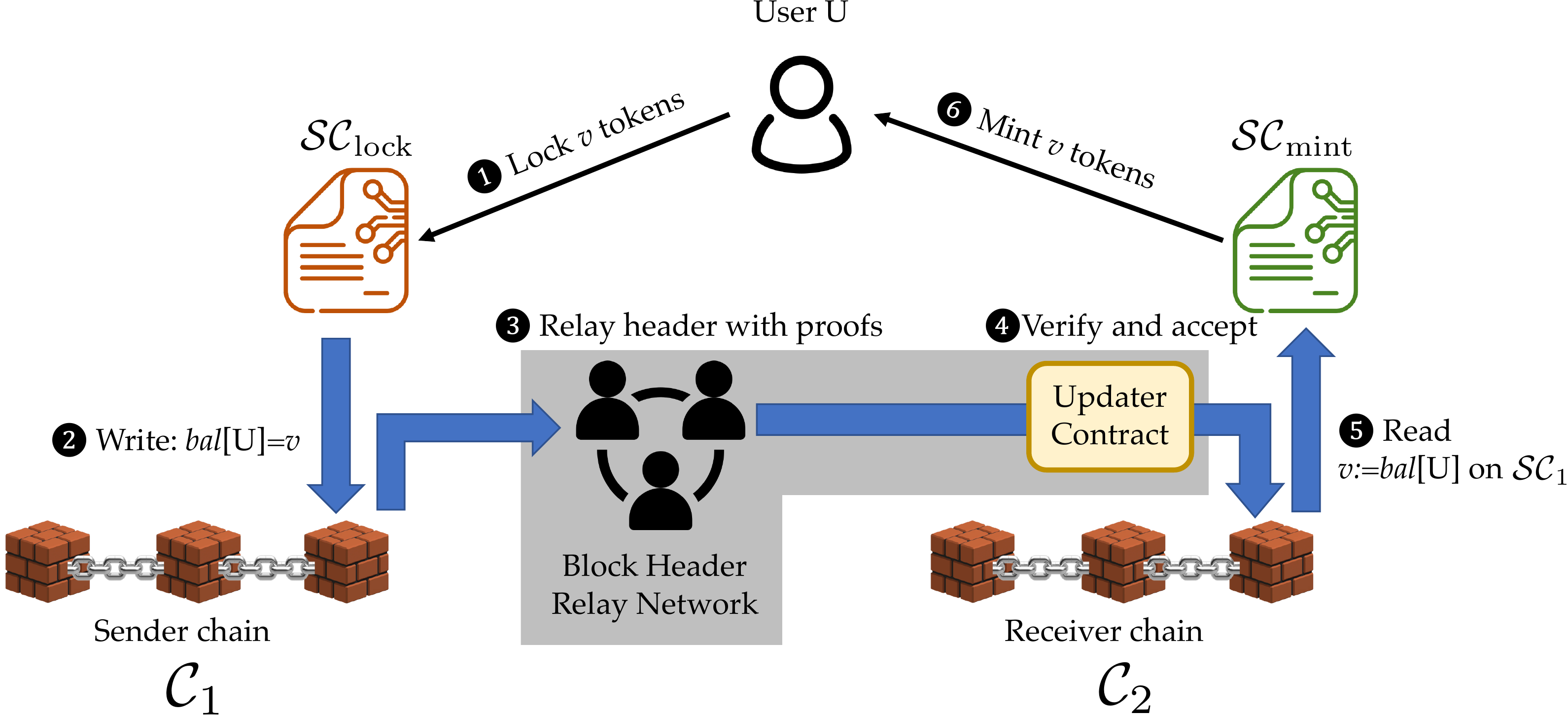}
    \caption{The design of \systemname illustrated with the example of cross-chain token transfer. The components in shade belongs to \systemname. For clarity we only show one direction of the bridge and the opposite direction is symmetric.}
    \label{fig:overview}
\end{figure*}

\subsection{Overview of \systemname design}
\label{motivating_example}

To make it easy for different applications to integrate with \systemname, we adopt a modular design where we separate application-specific logic (e.g., verifying smart contract states) from  the core bridge functionality (i.e., relaying block headers).

\Cref{fig:overview} shows the architecture and workflow of \systemname.
The core bridge functionality is provided by a {\bf block header relay network} (trusted only for liveness) that relays block headers of $\chainone$ along with correctness proofs, and an {\bf updater contract} on $\chaintwo$ that verifies and accepts proofs submitted by relay nodes. 
The updater contract maintains a list of recent block headers, and updates it properly after verifying proofs submitted by relay nodes; it exposes a simple and application-agnostic API, from which application smart contracts can obtain the latest block headers of the sender blockchain and build application-specific logic on top of it. 

Applications relying on \systemname will typically deploy a pair of contracts, a sender contract and a receiver contract on $\chainone$ and $\chaintwo$, respectively. We refer to them collectively as application contracts or relying contracts.
The receiver contract can call the updater contract to obtain block headers of $\chainone$, based on which they can perform application specific tasks. 
Depending on the application, receiver contracts might also need a user or a third party to provide application-specific proofs, such as Merkle proofs for smart contract states.

\newcommand{\contractlock}{\mathcal{SC}_\text{lock}}
\newcommand{\contractmint}{\mathcal{SC}_\text{mint}}

As an example, \cref{fig:overview} shows the workflow of {\em cross-chain token transfer}, a common use case of bridges, facilitated by \systemname.
Suppose a user $\user$ wants to trade assets (tokens) she owns on blockchain $\chainone$ in an exchange residing on another blockchain $\chaintwo$ (presumably because $\chaintwo$ charges lower fees or has better liquidity), she needs to move her funds from $\chainone$ to $\chaintwo$.
A pair of smart contracts $\contractlock$  and $\contractmint$ are deployed on blockchains $\chainone$ and $\chaintwo$ respectively.
To move the funds, the user
locks $\vdep$ tokens in $\contractlock$ (Step \Circled{1}  in~\cref{fig:overview}) and then requests  $\vdep$ tokens to be issued by  $\contractmint$.
To ensure solvency, $\contractmint$ should only issue new tokens if and only if the user has locked tokens on  $\chainone$.
This requires $\contractmint$ to read the states of $\contractlock$ (the balance of $\user$, updated in step \Circled{2}) from a different blockchain, which it cannot do directly.
\systemname enables this by relaying the block headers of $\chainone$ to $\chaintwo$ along with proofs (step \Circled{3} and \Circled{4}).
$\contractmint$ can retrieve the block headers from the smart contract frontend (the updater contract), check that the balance of user $\user$ is indeed $\vdep$ (step \Circled{5}), and only then mint $\vdep$ tokens (Step \Circled{6}).

Besides cross-chain token transfer, \systemname can also  enable various other applications such as cross-chain collateralized loans, general message passing, etc. We present three use cases in~\cref{section:applications}.

\newtheorem{protocol}{Protocol}

\subsection{Protocol detail}
\label{sec:system}

Having presented the overview, in this section, we specify the protocol in more detail.

\subsubsection{Security and system model}

\newcommand{\bridgeprotocol}{\mathcal{BR}[\contractone \leftrightarrow \contracttwo]}
\newcommand{\bridgeprotocolonedir}{\mathcal{BR}[\contractone\to\contracttwo]}

For the purpose of modeling bridges, we model a blockchain $\mathcal{C}$ as a block-number-indexed key-value store, denoted as $\mathcal{C}[t]:\mathcal{K}\to \mathcal{V}$ where $t$ is the block number, $\mathcal{K}$ and $\mathcal{V}$ are key and value spaces respectively.
In Ethereum, for example, $\mathcal{V}=\{0,1\}^{256}$ and keys are the concatenation of a smart contract identifier $\mathcal{SC}$ and a per-smart-contract storage address $K$.
For a given contract $\mathcal{SC}$, we denote the value stored at address $K$ at block number $t$ as $\mathcal{SC}[t, K]$, and we call $\mathcal{SC}[t, \cdot]$ the {\em state} of $\mathcal{SC}$ at block number $t$.
Again, for ease of exposition, we focus on the direction from $\contractone$ to $\contracttwo$, denoted as $\bridgeprotocolonedir$.

\paragraph{Functional and security goals.}
We require the bridge $\bridgeprotocolonedir$ to reflect states of $\contractone$ correctly and timely:

\begin{enumerate}[leftmargin=*]
    \item {\bf Correctness}: For all $t, K$, $\contracttwo$ accepts a wrong state $V \neq \contractone[t, K]$ with negligible probability.
    \item {\bf Liveness}: Suppose $\contracttwo$ needs to verify $\contractone$'s state at $(t, K)$, the bridge will provide necessary information eventually.
\end{enumerate}

\paragraph{Security assumptions.} For correctness, \systemname does not introduce extra trust assumptions besides those made by the underlying blockchains.
Namely, we assume both the sender blockchain and the receiver blockchain are consistent and live (\cref{sec:background}), \newcontent{and the sender chain has a light client protocol to enable fast block header verification.}
For both properties, we assume there is at least one honest node in the relay network, and that the \snark{} used is sound.

\subsubsection{Construction of \systemname}
\label{subsec:construction}

As described in~\cref{sec:overview}, a bridge $\bridgeprotocolonedir$ consists of three components: 
a block header relay network, a updater contract, and one or more application contracts.
Below we specify the protocols for each component.

\paragraph{Block header relay network.} 
We present the formal protocol of block header relay network in Protocol~\ref{protocol:brn}.

\begin{ourprotocol}
\caption{Block header relay network}\label{protocol:brn}
\begin{algorithmic}

\Procedure{RelayNextHeader}{$\lcstate_{r-1}, \blockheader_{r-1}$}
    \State Contact $k$ different full nodes to get the block headers following $\blockheader_{r-1}$, namely $\blockheader_r$.
    \State Generate a ZKP $\pi$ proving $$\lightclientcircuit(\lcstate_{r-1},\blockheader_{r-1},\blockheader_r) \to \true.$$
    \State Send $(\pi, \blockheader_{r}, \blockheader_{r-1})$ to the \updater contract.
\EndProcedure
\end{algorithmic}
\end{ourprotocol}

Nodes in the block header relay network run \textsf{RelayNextHeader} with the current state of the updater contract ($\lcstate_{r-1}, \blockheader_{r-1}$) as input.
The exact definition of $\lcstate_{r-1}$ is specific to light client protocols (see~\cite{chatzigiannis2021sok} for a survey).
The relay node then connects to full nodes in $\chainone$ and gets the block header $\blockheader_{r}$ following $\blockheader_{r-1}$. 
The relay node generates a ZKP $\pi$ showing the correctness of  $\blockheader_{r}$, by essentially proving that $\blockheader_{r}$ is accepted by a light client of $\chainone$ after block $\blockheader_{r-1}$.
It then sends $(\pi, \blockheader_{r})$ to the updater contract on $\chaintwo$.
To avoid the wasted proof time due to collision (note that when multiple relay nodes send at the same time, only one proof can be accepted), relay nodes can coordinate using standard techniques (e.g., to send in a round robin fashion).
While any zero-knowledge proofs protocol could be used, our highly optimized one will be presented later in~\cref{sec:dist}.

To incentivize block header relay nodes, provers may be rewarded with fees after validating their proofs. We leave incentive design for future work.
A prerequisite of any incentive scheme is unstealability~\cite{hyperproofs}, i.e., the guarantee that malicious nodes cannot steal others' proofs.
To this end, provers will embed their identifiers (public keys) in proofs, e.g., as input to the hash function in the Fiat-Shamir heuristic~\cite{fiat1986prove}.

We note that this design relies on the security of the light client verifier of the sender chain.
For example, the light client verifier must reject a valid block header
that may eventually become orphaned and not part of the sender chain.

\paragraph{The updater contract.} The protocol for the updater contract is specified in Protocol~\ref{protocol:updater}.

\newcommand{\headerList}{{\textsf{headerDAG}}}

\begin{ourprotocol}
\caption{The \updater contract}\label{protocol:updater}
\begin{algorithmic}
\State $\headerList := \emptyset$ \Comment{DAG of headers}
\State $\lcstate := \bot$ \Comment{light client state}
\Procedure{HeaderUpdate}{$\pi, \blockheader_r, \blockheader_{r-1}$}
    \If {$\blockheader_{r-1} \not \in \headerList$} 
    \State return False \Comment{skip if parent block is not in the DAG}
    \EndIf
    \If {$\pi$ verifies against $\lcstate,\blockheader_{r-1},\blockheader_{r}$}
    \State Update $\lcstate$ according to the light client protocol. 
    \State Insert $\blockheader_{r}$ into $\headerList$.
    \EndIf
\EndProcedure
\Procedure{GetHeader}{$t$} \Comment{$t$ is a unique identifier to a block header}
    \If {$t \not \in \headerList$} 
    \State return $\bot$ \Comment{tell the caller to wait}
    \Else
    \State return $\headerList[t], \lcstate$ \Comment{The $\lcstate$ will help users to determine if $t$ is on a fork.}
    \EndIf
\EndProcedure
\end{algorithmic}
\end{ourprotocol}

The updater contract maintains the light client's internal state including a list of block headers of $\chainone$ in $\headerList$.
It has two publicly exposed functions.
The \textsf{HeaderUpdate} function can be invoked by any block header relay node, providing supposedly the next block header and a proof as input. If the proof verifies against the current light client state $\lcstate$ and $\blockheader_{r-1}$, \newcontent{the contract will do further light-client checks, and then the state will be updated accordingly.}
Since the caller of this function must pay a fee, DoS attacks are naturally prevented.

The \textsf{GetHeader} function can be called by receiver contracts to get the block header at height $t$.
Receiver contracts can use the obtained block header to finish application-specific verification, potentially with the help of a user or some third party.

\paragraph{Application contracts.} 
\systemname has a modular design in that the updater contract is application-agnostic.
Therefore in $\bridgeprotocolonedir$, it is up to the application contracts $\contractone$ and $\contracttwo$ to decide what the information to bridge is.
Generally, proving that $\contractone[t, K]=V$ is straightforward: $\contracttwo$ can request for a Merkle proof for the leaf of the state Trie Tree (at block number $t$) corresponding to address $K$.
The receiver contract can obtain $\blockheader_t$ from the updater contract by calling the function \textsf{GetHeader($t$)}. 
Then it can verify $\contractone[t, K]=V$ against the Merkle root in $\blockheader_t$.
Required Merkle proofs are application-specific, and are typically provided by the users of $\contracttwo$, some third party, or the developer/maintainer of $\contracttwo$.

\paragraph{Security arguments.} The security of \systemname is stated in the following theorem.

\begin{theorem}\label{thm:bridge} The bridge $\bridgeprotocolonedir$ implemented by protocols \ref{protocol:brn} and \ref{protocol:updater} satisfies both consistency and liveness, assuming the following holds:
\begin{enumerate}
    \item there is at least one honest node in the block header relay network;
    \item the sender chain is consistent and live; 
    \item the sender chain has a light-client verifier as in Def.\@~\ref{def:lightclient}; and
    \item the succinct proof system is sound. 
\end{enumerate}
\end{theorem}

\begin{hproof}

To prove the consistency of DAG, we first need to convert the DAG into a list of blocks to match the definition of blockchain consistency. We define an algorithm $\longestchainfinder: \mathsf{DAG}\rightarrow \mathsf{List}$ such that given a DAG, the algorithm will output a list $\mathsf{MainChain}$ representing the main chain. For example, if the sender chain is Ethereum, the algorithm $\longestchainfinder$ will first calculate the path with the maximum total difficulty in the DAG represented by $\mathsf{L}$, and then output $\mathsf{MainChain} := \mathsf{L}[:-K]$. Here $K$ is a security parameter. By assumption $1$ and $2$, there will be an honest node in our system running either a full node or a light node, which will be consistent with the sender chain. Also,
according to assumption $1$, at least one prover node is honestly proving the light client execution. By assumption $4$ that the proof system is sound, the updater contract will correctly verify the light-client state.
We argue that the updater contract is correctly running the light-client protocol. Therefore, by the consistency of the light-client protocol, $\mathsf{MainChain}$ will be consistent with any other honest node.

The liveness of our protocol directly follows from the liveness of $\chainone$ and its light client protocol.
\end{hproof}

\subsection{Application use cases}
\label{section:applications}

In this section, we present three examples of applications that \systemname can support.

\newcommand{\TX}{\textsf{TX}}
\newcommand{\blockOne}{B_t}

\paragraph{Transaction inclusion: a building block.}
A common building block of cross-chain applications is to verify transaction inclusion on another blockchain.
Specifically, the goal is to enable a receiver contract $\contracttwo$ on $\chaintwo$ to verify that a given transaction $\trx$ has been included in a block $\blockOne$ on $\chainone$ at height $t$.
To do so, the receiver contract $\contracttwo$ needs a user or a third-party service to provide the Merkle proof for $\trx$ in $\blockOne$.
Then, $\contracttwo$ will call the updater contract to retrieve the block header of $\chainone$ at height $t$, and then verify the provided Merkle proof against the Merkle root contained in the header.

Next, we will present three use cases that extend the building block above.

\paragraph{1. Message passing and data sharing.} \label{app1}
Cross-chain message passing is another common building block useful for, e.g., sharing off-chain data cross blockchains.

Message passing can be realized as a simple extension of transaction inclusion, by embedding the message in a transaction.
Specifically, to pass a message $m$ from $\chainone$ to $\chaintwo$, a user can embed $m$ in a transaction $\trx_m$, send $\trx_m$ to $\chainone$, and then execute the above transaction inclusion proof.

\paragraph{2. Cross-chain assets transfer/swap.}
Bridging native assets is a common use case with growing demand. In this application, users can stake a certain amount of token $T_A$ on the sender blockchain $\chainone$, and get the same amount of token $T_A$ (for native assets transfer if eligible) or a certain amount of token $T_B$ of approximately the same value (for native assets swap) on the receiver blockchain $\chaintwo$. 
With the help of the transaction inclusion proof, native assets transfer/swap can be achieved, as illustrated at a high level in~\cref{motivating_example}. Here we specify the protocol in more detail.

To set up, the developers will deploy a lock contract $\contractlock$ on  $\chainone$ and a mint contract $\contractmint$ on  $\chaintwo$.
For a user who wants to exchange $n_A$ of token $T_A$ for an equal value in token $T_B$, she will first send a transaction $\trx_\text{lock}$ that transfers $n_A$ of token $T_A$ to $\contractlock$, along with an address $addr_{\chaintwo}$ to receive token $T_B$ on $\chaintwo$.  
After $\trx_\text{lock}$ is confirmed in a block $B$, the user will send a transaction $\trx_{\text{mint}}$ to $\contractmint$, including sufficient information to verify the inclusion of $\trx_{\text{lock}}$.
Based on information in $\trx_{\text{mint}}$, $\contractmint$ will verify that $\trx_\text{lock}$ has been included on $\chainone$, and transfer the corresponding $T_B$ tokens to the address $addr_{\chaintwo}$ specified in $\trx_{\text{lock}}$.
Finally, $\contractmint$ will mark $\trx_\text{lock}$ as minted to conclude the transfer.

\paragraph{3. Interoperations for NFTs.}
In the application of Non-fungible Token (NFT) interoperations, users always lock/stake the NFT on the sender blockchain, and get minted NFT or NFT derivatives on the receiver blockchain. By designing the NFT derivatives, the cross-chain protocol can separate the ownership and utility of an NFT on two blockchain systems, thus supporting locking the ownership of the NFT on the sender blockchain and getting the utility on the receiver blockchain.

\subsection{Efficient Proof Systems for \systemname}

The most computationally demanding part of \systemname is the zero-knowledge
proofs generation that relay nodes must do for every block.
So far we have abstracted away the detail of proof generation, which we will address in \cref{sec:dist,sec:composition}.
Here, we present an overview of our solution.

For Proof-of-Stake chains, the proofs involve verifying hundreds of signatures.
A major source of overhead is field transformation between different elliptic curves when the sender and receiver chains use different cryptography implementation, which is quite common in practice.
For example, Cosmos uses EdDSA on Curve25519 whereas Ethereum natively supports a different curve BN254.
The circuit for verifying a single Cosmos signature in the field supported by Ethereum involves around 2 million gates, thus verifying a block (typically containing 32 signatures) will involve over 64 million gates, which is too big for existing zero-knowledge proofs schemes. 

To make \systemname practical, we propose two ideas.

\paragraph{Reducing proof time with \devirgo}
We observe that the ZKP circuit for verifying multiple signatures is composed of multiple copies of one sub-circuit.
Our first idea is to take advantage of this special structure and distribute proof generation across multiple servers.
We propose a novel {\em distributed ZKP protocol} dubbed \devirgo, which carefully parallelizes the Virgo~\cite{virgo} protocol, one of the fastest ZKP systems (in terms of prover time) without a trusted setup.
With \devirgo, we can accelerate proof generation in \systemname with perfect linear scalability.
We will dive into the detail of \devirgo in~\cref{sec:dist}.

\paragraph{Reducing on-chain cost by recursive verification.}
While verifying \devirgo proofs on ordinary CPUs is very efficient, on-chain verification is still costly. 
To further reduce the on-chain verification cost (computation and storage), we use {\em recursive verification}: the prover recursively proves the correctness of a (potentially large) Virgo proof using a smart-contract-friendly zero-knowledge protocol to get a small and verifier-efficient proof.
At a high level, we trade slightly increased proof generation time for much reduced on-chain verification cost: the proof size reduces from $200+$KB to $131$ bytes, and the required computation reduces from infeasible amount of gas to $210$K gas. We will present more detail of recursive verification in Section~\ref{sec:composition}.

\section{Distributed proof generation}\label{sec:dist}

As observed previously, the opportunity for fast prover time stems from the fact that the circuit for verifying $N$ signatures consists of $N$ copies of identical sub-circuits. This type of circuits is called data-parallel~\cite{justinnote}. The advantage of data-parallel circuits is that there is no connection among different sub-copies. Therefore, each copy can be handled separately. We consider accelerating the proof generation on such huge circuits by dealing with each sub-circuit in parallel. In this section, we propose a distributed \snark{} protocol on data-parallel circuits. 

There are many zero knowledge proofs protocols~\cite{virgo, libra, spartan, hyrax, libsnark, aurora, ligero, libstark, virgoplus, gabizon2019plonk, fractal} supporting our computation. We choose Virgo as the underlying ZKP protocols for two reasons: 1. Virgo does not need a trusted setup and is plausibly post-quantum secure. 2. Virgo is one of the fastest protocols with succinct verification time and succinct proof size for problems in large scale. We present a new distributed version of Virgo for data-parallel arithmetic circuits achieving optimal scalability without any overhead on the proof size. Specifically, our protocol of \devirgo on data-parallel circuits with $\numProver$ copies using $\numProver$ parallel machines is $\numProver$ times faster than the original Virgo while the proof size remains the same. Our scheme is of independent interest and is possible to be used in other Virgo-based systems to improve the efficiency.

We provide the overall description of \devirgo as follows. Suppose the prover has $\numProver$ machines in total, labeled from $\P_0$ to $\P_{\numProver - 1}$. Assume $\P_0$ is the master node while other machines are ordinary nodes. Assume $\V$ is the verifier. Given a data-parallel arithmetic circuit consisting of $\numProver$ identical structures, the na\"ive algorithm of the distributed Virgo is to assign each sub-circuit to a separate node. Then each node runs Virgo to generate the proof separately. The concatenation of $N$ proofs is the final proof. Unfortunately, the proof size in this naive algorithm scales linearly in the number of sub-circuits, which can be prohibitively large for data-parallel circuits with many sub-copies. To address the problem, our approach removes the additional factor of $N$ in the proof size by aggregating messages and proofs among distributed machines. Specifically, the original protocol of Virgo consists of two major building blocks. One is the GKR protocol~\cite{GKR}, which consists of $d$ sumcheck protocols~\cite{sumcheck} for a circuit of depth $d$. The other is the polynomial commitment (PC) scheme. We design distributed schemes for each of the sumcheck and the polynomial commitment (PC). In our distributed sumcheck protocol, a master node $\P_0$ aggregates messages from all machines, then sends the aggregated message to $\V$ in every round, instead of sending messages from all machines directly to $\V$. Our protocol for distributed sumcheck has exactly the same proof size as the original sumcheck protocol, thus saving a factor $N$ over the na\"ive distributed protocol. Additionally, in our distributed PC protocol,  we optimize the commitment phase and make $\P_0$ aggregate $N$ commitments into one instead of sending $N$ commitments directly to $\V$. During the opening phase, the proof can also be aggregated, which improves the proof size by a logarithmic factor in the size of the polynomial. 

We present preliminaries in Section~\ref{sec:distpreliminary}, the detail of the distributed sumcheck protocol in Section~\ref{subsec:distsumcheck} and the detail of the distributed PC protocol in Section~\ref{subsec:distpc}. We combine them all together to build \devirgo in Section~\ref{subsec:distcombine}.

\subsection{Preliminaries}\label{sec:distpreliminary}
\paragraph{Multi-linear extension/polynomial. }
	Let $V:\{0, 1\}^\ell \rightarrow \mathbb{F}$ be a function. The \textit{multi-linear extension/polynomial} of $V$ is the unique polynomial $\tilde{V}: \mathbb{F}^\ell \rightarrow \mathbb{F}$ such that $\tilde{V}(\vecx) = V(\vecx)$ for all $\vecx\in\{0,1\}^\ell$. $\tilde{V}$ can be expressed as:
	$$\tilde{V}(\vecx)=\sum\nolimits_{\vecb\in\{0,1\}^\ell}\prod\nolimits_{i=1}^{\ell}((1-x_i)(1-b_i)+x_ib_i)) \cdot V(\vecb), $$
	where $b_i$ is $i$-th bit of b.

\paragraph{Identity function. }
	Let $\beta:\{0, 1\}^\ell \times \{0, 1\}^\ell \rightarrow \binary$ be the identity function such that $\beta(\vecx, \vecy) = 1$ if $\vecx = \vecy$, and $\beta(\vecx, \vecy) = 0$ otherwise. Suppose $\tbeta$ is the multilinear extension of $\beta$. Then $\tbeta$ can be expressed as:  
	$\tbeta(\vecx, \vecy) = \prod\nolimits_{i=1}^\ell ((1-x_i)(1-y_i) + x_iy_i)$.

\subsection{Distributed sumcheck}\label{subsec:distsumcheck}
\renewcommand{\P}{\mathcal{P}}
\renewcommand{\V}{\mathcal{V}}
\paragraph{Background: the sumcheck protocol.}The sumcheck problem is to sum a multivariate polynomial $f: \F^{\ell} \rightarrow \F$ over all binary inputs: $\sum_{b_1, \cdots, b_{\ell}\in\{0, 1\}}$ $f(b_1, \cdots, b_\ell)$. The sumcheck protocol allows the prover $\P$ to convince the verifier $\V$ that the summation is $H$ via a sequence of interactions, and the formal protocol is presented in Protocol~\ref{prot::sumcheck} in Appendix~\ref{sec:sumcheck}.
The high-level idea of the sumcheck protocol is to divide the verification into $\ell$ rounds. In each round, the prover only sends a univariate polynomial to the verifier. The verifier checks the correctness of the polynomial by a single equation. Then this variable will be replaced by a random point sampled by the verifier. As there are totally $\ell$ variables in $f$, after $\ell$ rounds, the claim about the summation will be reduced to a claim about $f$ on a random vector $\vecr$. Given the oracle access to $f$ on a random vector, the verifier can check the last claim. 

\paragraph{Background: the sumcheck equation in the GKR protocol.} In the GKR protocol working for a layered arithmetic circuit, both parties build a sumcheck equation to describe wire connections between the $i$-th layer and the $(i+1)$-th layer. Without loss of generality, we suppose there are $2^\ell$ gates in each layer. We define a polynomial $V_i: \binary^{\ell} \rightarrow \F$ such that $V_i(\vecb)$ represents the value of gate $\vecb$ in layer $i$, where $\vecb$ is the binary representation of integer $b$. We use $\tV_i$ as the multi-linear polynomial of $V_i$. Then we can write a sumcheck equation 
\begin{align}\label{eq:simpleGKR}
\tV_i(\vecg) = \sum\limits_{\vecx \in \binary^\ell}f(\vecx, \tV_{i+1}(\vecx)), 
\end{align}
where $f$ is some polynomial from $\F^\ell$ to $\F$ and $\vecg$ is a random vector in $\F^\ell$. By invoking the sumcheck protocol, the prover reduces a claim about the $i$-th layer to a claim about the $(i+1)$-th layer. Suppose the circuit depth is $d$, after running $d$ sumcheck protocols, the prover reduces the claim about the output layer to the input layer, which the verifier itself can verify. Due to the space limitation, we present the formal GKR protocol in Protocol~\ref{prot:GKR} in Appendix~\ref{sec:gkr}. 

We treat Equation~\ref{eq:simpleGKR} as the sumcheck equation in the GKR protocol and give the complexity of the sumcheck protocol running on Equation~\ref{eq:simpleGKR} as follows.

\paragraph{Complexity of the sumcheck protocol. }For the multivariate polynomial of $f$ defined in Equation~\ref{eq:simpleGKR}, the prover time in Protocol~\ref{prot::sumcheck} is  $O(2^\ell)$. The proof size is $O(\ell)$ and the verifier time is $O(\ell)$. 

In the setting of data-parallel circuits, we distribute the sumcheck polynomial $f$ among parallel machines. Suppose the data-parallel circuit $\circuit$ consists of $N$ identical sub-circuits of $\circuit_0, \cdots, \circuit_{N-1}$ and $\numProver = 2^n$ for some integer $n$ without loss of generality. The polynomial $f: \F^\ell \rightarrow \F$ is defined on $C$ by Equation~\ref{eq:simpleGKR}. 

The idea of our distributed sumcheck protocol is to treat each sub-copy as a new circuit as there is no wiring connections across different sub-circuits. We define polynomials of $f^{(0)}, \cdots, f^{(N-1)}$ on $\circuit_0, \cdots, \circuit_{N-1}: \F^{\ell - n} \rightarrow \F$ respectively by Equation~\ref{eq:simpleGKR} in the GKR protocol, which have the same form as $f$ defined on $C$. The na\"ive approach is running the sumcheck protocol on these polynomials separately. As there are $N$ proofs in total and each size is $O(\ell - n)$, the total proof size will be $O(N(\ell-n))$. To reduce the proof size back to $\ell$, the prover needs to aggregate $N$ proofs to generate a single proof on $f$. We observe that the sumcheck protocol on data-parallel circuits satisfies $f^{(i)}(\vecx)$ $ = $ $f(\vecx, \veci)$. 
As shown in Protocol~\ref{prot::sumcheck}, the protocol proceeds for $\ell$ variables round by round. We first run the sumcheck protocol on variables that are irrelevant to the index of sub-copies in the circuit. In the first $(\ell - n)$ rounds, each prover $\P_i$ generates the univariate polynomial of $f^{(i)}_j(x_j)$ for $f^{(i)}(\vecx)$ and sends it to $\P_0$. $\P_0$ constructs the univariate polynomial for $f_j(x_j)$ by summing $f^{(i)}_j(x_j)$ altogether since $f_j(x_j) = \sum\limits_{i=0}^Nf^{(i)}_j(x_j)$, and sends $f_j(x_j)$ to $\V$ in the $j$-th round. The aggregation among parallel machines reduces the proof size to constant in each round. Hence the final proof size is only $O(\ell)$. A similar approach has appeared in~\cite{wahby2016verifiable}. The main focus of~\cite{wahby2016verifiable} was improving the prover time of the sumcheck protocol in the GKR protocol to $O(2^\ell(\ell - n))$ for data-parallel circuits, which was later subsumed by~\cite{libra} with a prover running in $O(2^\ell)$ time. Instead, our scheme is focused on improving the prover time by $N$ times with distributed computing on $N$ machines without any overhead on the proof size.

With this idea in mind, we rewrite the sumcheck equation on $f$ as follows. 
$$H = \sum_{\vecb\in\{0, 1\}^\ell}f(\vecb) = \sum_{i=0}^{N-1}\sum_{\vecb\in \{0, 1\}^{\ell-n}}f^{(i)}(\vecb).$$
Then we divide the original sumcheck protocol on $f$ into 3 phases naturally in the setting of distributed computing. We present the formal protocol of distributed sumcheck in Protocol~\ref{prot::distsumcheck} in Appendix~\ref{sec:distsumcheck}.
\begin{enumerate}
    \item From round 1 to round $(\ell - n)$ (step~\ref{step: distsumcheck1} in Protocol~\ref{prot::distsumcheck}), $\P_i$ runs the sumcheck protocol on $f^{(i)}$ and sends the univariate polynomial to $\P_0$. After receiving all univariate polynomials from other machines, $\P_0$ aggregates these univariate polynomials by summing them together and sends the aggregated univariate polynomial to the verifier. When $\P_0$ receives a random query from the verifier, $\P_0$ relays the random challenge to all nodes as the random query of the current round. 
    \item In round $(\ell - n)$ (step~\ref{step: distsumcheck2} in Protocol~\ref{prot::distsumcheck}), the polynomials of $f^{(0)}, \cdots, f^{(N-1)}$ have been condensed to one evaluation on a random vector  $\vecr\in\F^{\ell-n}$. $\P_0$ uses these $N$ points as an array to construct the multi-linear polynomial $f': \F^n \rightarrow \F$ such that $f'(\vecx) = f(\vecr, \vecx[1:n])$.\footnote{The approach can extend to the product of two multi-linear polynomials, which matches the case in Virgo.}
    \item After round $(\ell - n)$ (step~\ref{step: distsumcheck3} in Protocol~\ref{prot::distsumcheck}), $\P_0$ continues to run the sumcheck protocol on $f'$ with $\V$ in last $n$ rounds. 
\end{enumerate}
In this way, the computation of $\P_i$ is equivalent to running the sumcheck protocol in Virgo on $\circuit_i$. It accelerates the sumcheck protocol in Virgo by $\numProver$ times without any overhead on the proof size using $\numProver$ distributed machines, which is optimal for distributed algorithms both in asymptotic complexity and in practice. We give the complexity of Protocol~\ref{prot::distsumcheck} in the following.

\paragraph{Complexity of the distributed sumcheck protocol.}
For the multivariate polynomial of $f$ defined in Equation~\ref{eq:simpleGKR}, The total prover work is $O(2^\ell)$ while the prover work for each machine is $O(\frac{2^\ell}{N})$. The communication between $\numProver$ machines is $O(N\ell)$. The proof size and the verifier time are both $O(\ell)$.
\renewcommand{\P}{\mathcal{P}}
\renewcommand{\V}{\mathcal{V}}
\renewcommand{\F}{\mathbb{F}}

\subsection{Distributed polynomial commitment}\label{subsec:distpc}
In the last step of the sumcheck phase, the prover needs to prove to the verifier $y = f(r_1, \cdots, r_\ell)$ for some value $y$. In Virgo, The prover convinces $\V$ of the evaluation by invoking the PC scheme. We present the PC scheme in Virgo and the complexity of the scheme in the following.

\paragraph{Background: the polynomial commitment in Virgo. }\label{def:pc}
Let $\mathcal{F}$ be a family of $\ell$-variate multi-linear polynomial over $\F$. Let $\H$, $\L$ be two disjoint multiplicative subgroups of $\F$ such that $|\H| = 2^\ell$ and $|\L| = \rho|\H|$, where $\rho$ is a power of 2. The polynomial commitment (PC) in Virgo for $f \in \mathcal{F}$ and $\vecr \in\F^\ell$ consists of the following algorithms: 
\begin{itemize}
    \item $\pp \leftarrow \PC.\KeyGen(1^\lambda)$: Given the security parameter $\lambda$, the algorithm samples a collision resistant hash function from a hash family as $\pp$.
    \item $\com_f \leftarrow \PC.\Commit(f, \pp)$: Given a multi-linear polynomial $f$, the prover treats $2^\ell$ coefficients of $f$ as evaluations of a univariate polynomial $f_U$ on $\H$. The prover uses the inverse fast Fourier transform (IFFT) to compute $f_U$. Then the prover computes $\vecf_\L$ as evaluations of $f_U$ on $\L$ via the fast Fourier transform (FFT). Let $\com_f = \MT.\Commit(\vecf_\L)$. 

    \item $(y,\pi_f)\leftarrow \PC.\Open(f,\vecr, \pp)$: The prover computes $y = f(\vecr)$. \ignore{\color{blue} To validate the evaluation, the prover computes $\log |\L|$ polynomials of $q_1, \cdots, q_{\log |\L|}$ being size of $\frac{|\L|}{2}$, $\cdots$, $1$ depending on $q$. Then the prover commits all of these polynomial to get $\com_{q_1}$, $\cdots$, $\com_{q_{\log|\L|}}$. 
    Given $c =O(\lambda)$ random indexes $(k_1, \cdots, k_c)$, the prover computes $(\vecf_\L[k_1], \pi_{k_1}) = \MT.\Open(\vecf_\L, k_1)$, $\cdots$, $(\vecf_\L[k_c], \pi_{k_c}) $ $=$ $ \MT.\Open(\vecf_\L, k_c)$. Additionally, the prover also opens corresponding $c$ locations on $q_1$, $\cdots$, $q_{\log|\L|}$. Suppose the values on $q_1$, $\cdots$, $q_{\log|\L|}$ are $\vecv$ and the aggregation of the Merkle proof is $\pi_{agg}$. Let $\pi_f$ = $(\vecf_\L[k_1], \pi_{k_1}$, $\cdots$, $\vecf_\L[k_c], \pi_{k_c}, \vecv, \pi_{agg})$. } Given $c =O(\lambda)$ random indexes $(k_1, \cdots, k_c)$, the prover computes $(\vecf_\L[k_1], \pi_{k_1}) = \MT.\Open(\vecf_\L, k_1)$, $\cdots$, $(\vecf_\L[k_c], \pi_{k_c}) $ $=$ $ \MT.\Open(\vecf_\L, k_c)$. Let $\pi_f$ = $(\vecf_\L[k_1], \pi_{k_1}$, $\cdots$, $\vecf_\L[k_c], \pi_{k_c})$.\footnote{The prover also computes $\log |\L|$ polynomials of $f_1, \cdots, f_{\log|\L|}$ depending on $f$. But sizes of these polynomials are $\frac{|\L|}{2}, \cdots, 1$ respectively. The prover commits these polynomial and opens them on at most $c$ locations correspondingly. Our techniques on distributed commitment and opening can apply to these smaller polynomials easily. We omit the process for simplicity. It brings a logarithmic factor in the size of the polynomial on the proof size and the verification time.}
    
    \item $\{\accept,\reject\}\leftarrow\PC.\Verify(\com_f,\vecr,y,\pi_f,\pp)$: The verifier parses $\pi_f = (\vecq_\L[k_1], \pi_{k_1}, \cdots, \vecq_\L[k_c], \pi_{k_c})$, then checks that $\vecq_\L[k_1]$, $\cdots$, $\vecq_\L[k_c]$ are consistent with $y$ by a certain equation $p(f_\L[k_1]$, $\cdots$, $f_\L[k_c], y) = 0$, \footnote{$p$ also takes all openings on polynomials of $f_1, \cdots, f_{\log |\L|}$ (at most $c$ for each polynomial) as input, we omit them for simplicity. } and checks that $\vecf_\L[k_1]$, $\cdots$, $\vecf_\L[k_c]$ are consistent with $\com_f$ by $\MT.\Verify$ $(\pi_{k_1}, \vecf_\L[k_1]$, $\com_f)$, $\cdots$, $\MT.\Verify$ $(\pi_{k_c}, \vecf_\L[k_c], \com_f)$. If all checks pass, the verifier outputs \accept, otherwise the verifier outputs \reject.
\end{itemize}
\paragraph{Complexity of PC in Virgo. }The prover time is $O(\ell \cdot 2^\ell)$. The proof size is $O(\lambda\ell^2)$ and the verifier time is $O(\lambda\ell^2)$. 

In the setting of distributed PC,  $\P_i$ knows $f^{(i)}$. With the help of $\tbeta$ function, we have
\begin{align}\label{eq:random_f}
f(\vecr) = \sum_{i = 0}^{N-1}\tbeta(\vecr[\ell -n  + 1:\ell], \veci)f^{(i)}(\vecr[1:\ell - n]).
\end{align}
A straightforward way for distributed PC is that $\P_i$ runs the PC scheme on $f^{(i)}$ separately. In particular, $\P_i$ invokes $\PC.\Commit$ to commit $f^{(i)}$ in the beginning of the sumcheck protocol. In the last round, $\P_i$ runs $\PC.\Open$ to compute $f^{(i)}(\vecr[1:\ell - n])$ and sends the proof to $\V$. After receiving all $f^{(i)}(\vecr[1:\ell - n])$ from $\P_i$, $\V$ invokes $\PC.\Verify$ to validate $N$ polynomial commitments separately. Then $\V$ computes $\tbeta(\vecr[\ell -n  + 1:\ell], \veci)$ for each $i$.
Finally, $\V$ checks $f(\vecr) = \sum_{i = 0}^{N-1}\tbeta(\vecr[\ell -n  + 1:\ell], \veci)f^{(i)}(\vecr[1:\ell - n])$.

Although the aforementioned na\"ive distributed protocol achieves $O(2^\ell(\ell - n))$ in computation time for each machine, the total proof size is $O(\lambda N(\ell-n)^2)$ as the individual proof size for each $\P_i$ is $O(\lambda(\ell-n)^2)$. To reduce the proof size, we optimize the algorithm by aggregating $N$ commitments and $N$ proofs altogether. For simplicity, we assume $\rho = 1$ without loss of generality in the multi-linear polynomial commitment\footnote{In Virgo, $\rho = 32$ for security requirements. Our scheme can extend to $\rho = 32$ easily.}. We present the formal protocol of distributed PC in Protocol~\ref{prot::distpc} in Appendix~\ref{sec:distpcprotocolonly}.

The idea of our scheme is that each $\P_i$ exchanges data with other machines immediately after computing $\vecf^{(i)}_\L$ instead of invoking $\MT.\Commit$ on $\vecf^{(i)}_\L$ directly. The advantage of such arrangement is that the prover aggregates evaluation on the same index into one branch and can open them together by a single Merkle tree proof for this branch. As described in the polynomial commitment of Virgo, the prover needs to open $f_\L$ on some random indexes depending on $\vecr$ in $\PC.\Open$. As $\vecr$ is identical to each $f^{(i)}$, the prover would open each $f_\L^{(i)}$  at same indexes. If the prover aggregates $f_\L^{(i)}$ by the indexes, she can open $N$ values in one shot by providing only one Merkle tree path instead of na\"ively providing $N$ Merkle tree paths, which helps her to save the total proof size by a logarithmic factor in the size of the polynomial.

Specifically, $\P_i$ collects evaluations of $\vecf^{(0)}_\L[i+1]$, $\cdots$, $\vecf^{(N-1)}_\L[i+1]$ with identical index of $(i+1)$ in $\L$ from other machines (step~\ref{step:distpc1} and step~\ref{step:distpc2}). Then $\P_i$ invokes $\MT.\Commit$ to get a commitment, $com_{h^{(i)}}$, for these values, and submits $com_{h^{(i)}}$ to $\P_0$ (step~\ref{step:distpc3}). $\P_0$ invokes $\MT.\Commit$ on $com_{h^{(0)}}$, $\cdots$, $com_{h^{(N-1)}}$ to compute the aggregated commitment, $\com$, and $\P_0$ sends $\com$ to $\V$ (step~\ref{step:distpc4}). In the $\PC.\Open$ phase, given a random index $k_j$ from $\V$, $\P_0$ retrieves $\vecf^{(N-1)}_\L[k_j]$, $\cdots$, $\vecf^{(N-1)}_\L[k_j]$ from $\P_{k_j-1}$, computes $(\com_{h^{(k_j-1)}}, \pi_{k_j})$ = $\MT.\Open$ ($\com, k_j)$, and sends these messages to $\V$ (step~\ref{step:distpc5} and step~\ref{step:distpc6}). $\V$ can validate $N$ evaluations by invoking $\MT.\Verify$ only once (step~\ref{step:distpc7}). With this approach, we reduce the proof size to $O(\lambda(N + \ell^2))$. 

And the complexity of Protocol~\ref{prot::distpc} is shown in the following.

\paragraph{Complexity of distributed PC.}
Given that $f$ is a multi-linear polynomial with $\ell$ variables, the total communication among $N$ machines is $O(2^\ell)$. The total prover work is $O(2^\ell \cdot \ell)$ while the prover work for each device is $(\frac{2^\ell}{N}\cdot \ell)$. The proof size is $O(\lambda(N + \ell^2))$. The verification cost is $O(\lambda(N + \ell^2))$. 
\renewcommand{\P}{\mathcal{P}}
\renewcommand{\V}{\mathcal{V}}
\renewcommand{\F}{\mathbb{F}}

\subsection{Combining everything together}\label{subsec:distcombine}

In this section, we combine the distributed sumcheck and the distributed PC altogether to build \devirgo. 

For a data-parallel layered arithmetic circuit $C$ with $N$ copies and $d$ layers, following the workflow of Virgo in Protocol~\ref{prot:virgo} in Appendix~\ref{sec:virgo},  our distributed prover replaces $d$ sumcheck schemes in Virgo by $d$ distributed sumcheck schemes,  and replaces the PC scheme in Virgo by our distributed PC scheme to generate the proof. We present the formal protocol of \devirgo in Protocol~\ref{prot:distvirgo} in Appendix~\ref{sec:distvirgo}. And we have the theorem as follows.
\begin{theorem}\label{thm:main}
Protocol~\ref{prot:distvirgo} is an argument of knowledge satisfying the completeness and knowledge soundness in Definition~\ref{def::zkp} for the relation $C(\vecx, \vecw) = \vecone$, where $C$ consists of $N$ identical copies of $C_0, \cdots, C_{N-1}$.
\end{theorem}

\begin{hproof}
\paragraph{Completeness. }The completeness is straightforward. 

\paragraph{Knowledge soundness. }
deVirgo generates the same proof as Virgo for $d$ sumcheck protocols. So we only need to consider the knowledge soundness of distributed PC scheme. If the commitment of $f$ is inconsistent with the opening of $f(\vecr)$ in the distributed PC scheme, there must exist at least one $f^{(i)}(\vecr[1:\ell - n])$ being inconsistent with the commitment $f$ by Equation~\ref{eq:random_f}. Otherwise, when all $f^{(i)}(\vecr[1:\ell - n])$ are consistent with the commitment of $f$, $f(\vecr)$ must be consistent with the commitment of $f$. As shown in Protocol~\ref{prot::distpc}, $com_f$ is equivalent to $com_{f^{(i)}}$ with additional dummy messages in each element of the vector in the Merkle tree commitment. It does not affect the soundness of the PC in Virgo in the random oracle model~\cite{virgo, zhang2022polynomial}. The verifier outputs $0$ in the $\PC.\Verify$ phase with the probability of $(1-\negl(\lambda))$. Therefore, deVirgo still satisfies knowledge soundness.

The zero-knowledge property is not necessary as there is no private witness in the setting of zkbridge. However, we can achieve zero-knowledge for deVirgo by adding some hiding polynomials. Virgo uses the same method to achieve zero-knowledge.
\end{hproof}

Additionally, Fiore and Nitulescu~\cite{fiore2016security} introduced the notion of O-SNARK for SNARK over  authenticated data such as cryptographic signatures. Protocol~\ref{prot:distvirgo} is an O-SNARK for any oracle family, albeit in the random oracle model. To see this, Virgo relies on the construction of computationally sound proofs of Micali~\cite{Micali00} to achieve non-interactive proof and knowledge soundness in the random oracle model, which has been proven to be O-SNARK in~\cite{fiore2016security}. Hence Virgo is an O-SNARK, and so is deVirgo because deVirgo also relies on the same model. 

Protocol~\ref{prot:distvirgo} achieves optimal linear scalability on data-parallel circuits without significant overhead on the proof size. In particular, our protocol accelerates Virgo by $N$ times given $N$ distributed machines. Additionally, the proof size in our scheme is reduced by a factor of $N$ compared to the na\"ive solution of running each sub-copy of data-parallel circuits separately and generating $N$ proofs. The complexity of Protocol~\ref{prot:distvirgo} is shown in the following. 

\paragraph{Complexity of distributed Virgo. }Given a data-parallel layered arithmetic circuit $C$ with $N$ sub-copies, each having $d$ layers and $m$ inputs, the total prover work of Protocol~\ref{prot:distvirgo} is $O(|C| + Nm\log m)$. The prover work for a single machine is $O(|C|/N + m\log m)$, and the total communication among machines is $O(Nm + Nd\log |C|)$. The proof size is $O(d\log |C| + \lambda(N + \log^2 m))$. The verification cost is $O(d\log |C| + \lambda(N + \log^2 m))$.
\renewcommand{\P}{\mathcal{P}}
\renewcommand{\V}{\mathcal{V}}
\renewcommand{\F}{\mathbb{F}}

\section{Reducing proof size and verifier time}
\label{sec:composition}

Although \devirgo improves the prover time by orders of magnitude, we want to further reduce the cost of the verification time and the proof size. As mentioned in the above section, the circuit which validates over 100 signatures is giant due to non-compatible instructions on different curves across different blockchains. Additionally, Virgo's proof size, which is around 210KB for a circuit with 10 million gates, is large in practice. 

Thus we cannot post \devirgo's proof on-chain and validate the proof directly. Aiming at smaller proof size and simpler verification on-chain, we propose to further compress the proof by recursive proofs with two layers. Intuitively, for a large-scale statement  $(\vecx, \vecw) \in \R$ in Definition~\ref{def::zkp}, the prover generates the proof $\pi_1$ by a protocol with fast prover time in the first layer. If the length of $\pi_1$ is not as short as desired, then the prover can produce a shorter proof $\pi_2$ by invoking another protocol for $(\vecx, \pi_1) \in \R'$ in the second layer, where $\R'$ represents that $\pi_1$ is a valid proof for $(\vecx,\vecw) \in \R$. 
To shrink the proof size and simplify the verification as much as possible, we choose Groth16 as the second layer ZKP protocol since Groth16 has constant proof size and fast verification time. Moreover, the curve in Groth16 is natively supported by Ethereum, which is beneficial for saving on-chain cost on Ethereum. In our approach, the prover invokes \devirgo to generate $\pi_1$ on the initial circuit in the first layer. In the second layer, the prover invokes Groth16 to generate $\pi_2$ on the circuit implementing the verification algorithm of \devirgo where $|\pi_2| \ll |\pi_1|$.  The prover only needs to submit $\pi_2$ on-chain for verification. The recursion helps cross-chain bridges to reduce gas cost on blockchains because of simple verification on the compatible curve. The security of recursive proofs relies on random oracle assumption, which can be instantiated by a cryptographic hash function in practice~\cite{fractal}.

\begin{table}
\centering
  \begin{tabular}{ | m{3.5em} | m{2.3cm}| m{1.8cm} | m{1.8cm}| }
  \hline
  \# of sigs & Total circuit size &Circuit size for GKR part& Circuit size for PC part\\
  \hline
  1 & $1.2\times 10^{7}$ gates& $8.4\times 10^{6}$ gates & $3.3\times 10^{6}$ gates\\
  \hline
  4 &$1.2\times 10^{7}$ gates & $8.4\times 10^{6}$ gates & $4.0\times 10^{6}$ gates\\
  \hline
  32 & $1.3\times 10^{7}$ gates & $8.4\times 10^{6}$ gates & $4.7\times 10^{6}$ gates\\
  \hline
  128 & $1.4\times 10^{7}$ gates & $8.4\times 10^{6}$ gates & $5.4\times 10^{6}$ gates\\
  \hline
\end{tabular}
\vspace{.5cm}
\caption{The verification circuit size of \devirgo}

\label{table:circuit_size}
\end{table}

\paragraph{Performance gains.}
We use the signature validation circuit for Cosmos~\cite{cosmos} as an example to show concrete numbers of the verification circuit of \devirgo in Table~\ref{table:circuit_size}. 
We record the size of the whole verification circuit in the $2^{nd}$ column, the size for the GKR part in the $3^{rd}$ column, and the size for the PC part in the $4^{th}$ column, as the number of signatures in data-parallel circuits increases from 1 to 128 in the $1^{st}$ column. The number of gates in the $2^{nd}$ column equals the sum of numbers of gates in the $3^{rd}$ column and the $4^{th}$ column. As shown in Table~\ref{table:circuit_size}, although the data-parallel circuit size expands, the size for the sumcheck part in \devirgo's verification circuit does not change. That is because the verification for the GKR part is only based on the structure of the sub-circuit, which is identical among different copies. However, the size for the PC part in \devirgo's verification circuit up-scales sub-linearly in the number of copies due to the growth of the polynomial size. Even given 128 copies of the signature validation circuit, the bottleneck of \devirgo's verification circuit is the sumcheck part. Therefore, the recursive proof size and the recursive verification cost are independent of the number of signatures to validate in our instance. 
In addition, the prover time of Groth16 on the verification circuit of \devirgo is only 25\%  of the prover time of \devirgo in practice. Therefore, our recursive proof scheme reduces the on-chain proof verification cost from  $ \sim 8\times 10^7$ gas (an estimation) to less than $2.3\times 10^5$ gas.

\renewcommand{\P}{\mathcal{P}}
\renewcommand{\V}{\mathcal{V}}
\renewcommand{\F}{\mathbf{F}}

\section{Implementation and Evaluation}\label{sec:exp}

To demonstrate the practicality of \systemname, we implement a prototype from Cosmos~\cite{cosmos} (a PoS blockchain built on top of the Tendermint~\cite{kwon2014tendermint} protocol) to Ethereum, and from Ethereum to other EVM-compatible chains such as BSC. Supports for other blockchains can be similarly implemented with additional engineering effort, as long as they support light client protocols defined in Definition \ref{def:lightclient}. In this section, we discuss implementation detail, its performance, as well as operational cost.

The bridge from Cosmos to Ethereum is realized with the full blown \systemname protocol presented so far to achieve practical performance. In comparison, the direction from Ethereum to other EVM-compatible chains incurs much less overhead for proof generation and does not require \devirgo. Therefore, in what follows, we mainly focus on the direction from Cosmos to Ethereum.

\subsection{Implementation details}

The bridge from Cosmos to Ethereum consists of four components: a relayer that fetches Cosmos block headers and sends them to Ethereum (implemented in 300+ lines of Python), \devirgo (implemented in 10000+ lines of C++) for distributed proof generation, a handcrafted recursive verification circuit, and an updater contract on Ethereum (implemented in 600+ lines of Solidity). Our signature verification circuit is based on the optimized signature verification circuit \cite{eddsaCirc}. However, we use Gnark instead of Circom as in ~\cite{eddsaCirc} for better efficiency for proof generation.

\subsubsection{Generating correctness proofs.}
Relay nodes submit Cosmos block headers to the updater contract on Ethereum along with correctness proofs, which proves that the block is properly signed by the Cosmos validator committee appointed by the previous block. (In Cosmos a hash of the validator committee members is included in the previous block.)

In Cosmos, each block header contains about 128 EdDSA signatures (on Curve25519), Merkle roots for transactions and states, along with other metadata, where $32$ top signatures are required to achieve super-majority stakes. 
However, the most efficient curve supported by the Ethereum Virtual Machine (EVM) is BN254.
To verify Cosmos digital signatures in EVM, one must simulate Curve25519 on curve BN254, which will lead to large circuits.
Concretely, to verify a Cosmos block header (mainly, to verify about 32 signatures), we need about 64 million gates.
We implement \devirgo (\cref{sec:dist}) and recursive verification (\cref{sec:composition}) to accelerate proof generation and verification.

Moreover, in practical deployment, multiple relayers can form a pipeline to increase the throughput. Looking ahead, based on the evaluation results, our implementation can handle $1$ second block time in Cosmos with $120+$ capable relayers in the network.

For proof verification, we build an outer circuit that verifies Virgo proofs  and use Gnark~\cite{gnark} to generate the final Groth16 proof that can be efficiently verified by the updater contract on Ethereum.

\subsubsection{The updater contract.}
We implement the updater contract on Ethereum in Solidity that verifies Groth16 proofs and keeps a list of the Cosmos block headers in its persistent storage. The cost of verifying a Groth16 proof on-chain is less than $230K$ gas.

The updater contract exposes a simple API which takes block height as its input, and returns the corresponding block header.
The receiver contracts can then use the block header to complete application-specific verification.

\paragraph{Batching.} Instead of calling the updater contract on every new block header, we implemented {\em batching} where the updater contract stores Merkle roots of batches of $B$ consecutive block headers. The prover will first collect $B$ consecutive blocks, and then makes a unified proof for all $B$ blocks. The updater contract will only need to verify one proof for the batch of $B$ blocks.
After the verification, the updater contract checks the difficulty, stores the block headers, and updates the light-client state.
Storing one Merkle root every $B$ blocks also reduces storage cost. 
Thus $B$ can be set to balance user experience and cost: With a larger $B$, users need to wait longer, but the cost of running the system is lower.

We implement the aforementioned batched proof verification and show the experimental results in Section \ref{sec:exp_eval}. In addition, we propose a more complex batching optimization presented in Appendix \ref{app:optimisc} for further optimization.

With batching, the cost for storing  block headers and maintaining light-client states is amortized across $B$ blocks. The bulk of the cost incurred by the updater contract is SNARK proof verification, which is the focus of our evaluation below.

\subsection{Evaluation}\label{sec:exp_eval}

We evaluate the performance of \systemname (from Cosmos to Ethereum) from four aspects: proof generation time, proof generation communication cost, proof size, and on-chain verification cost.

\subsubsection{Experiment setup.}
We envision that a relayer node in \systemname will be deployed as a service in a managed network, therefore we evaluate \systemname in a data-center-like environment.
Specifically, we run all the experiments on {128} AWS EC2 c5.24xlarge instances with the Intel(R) Xeon(R) Platinum 8275CL CPU @ 3.00GHz and 192GB of RAM. 
Our implementation for the proof generation is parallelized with at most 128 machines. We report the average running time of 10 executions.
Whenever applicable, we report costs both in terms of running time and monetary expenses.

\begin{table*}[]
\begin{tabular}{|l|l|l|l|l|l|l|l|l|l|}
\hline
       & \multicolumn{3}{c|}{\bf Proof Gen. Time (seconds)}              & \multicolumn{2}{c|}{\bf Proof Gen. Comm. (GB)} & \multicolumn{2}{c|}{\bf Proof Size (Bytes)} & \multicolumn{2}{c|}{\bf On-chain Ver. Cost (gas)} \\
\hline
\# of sigs & \devirgo  & RV & total  & total   & per-machine  & w/o RV  & w/ RV  & w/o RV     & w/ RV    \\
\hline
 8 &   12.52  & 4.90 & 17.42  &  7.34 & 0.92  & 1946476 & 131 & 78M & 227K \\
\hline
 32 & 12.80 &  5.41 & 18.21 &  32.24  & 1.01 & 1952492  & 131 & 78M & 227K \\
\hline
 128  &  13.28  & 5.49 &  18.77 & 131.89 & 1.03  & 1958508 &  131 & 79M & 227K  \\
\hline
\end{tabular}
\vspace{.5cm}
\caption{Evaluation results. RV is the shorthand for recursive verification.\label{tab:eval}}
\end{table*}

\subsubsection{Proof generation time of \devirgo.}
We first evaluate the main cryptographic building block---\devirgo---and compare its performance with the original Virgo~\cite{virgo}.
The source code of the original Virgo is obtained at \url{https://github.com/sunblaze-ucb/Virgo}.
We run both protocols on the same circuit for correctness proofs, which mainly consists of $N$ invocation of EdDSA signature verification.

\begin{figure}[t]
    \centering
    \begin{subfigure}{.4\textwidth}
        \definecolor{colorA}{HTML}{e6194B}%
\definecolor{colorB}{HTML}{000000}
\definecolor{colorC}{HTML}{0000FF}
\definecolor{colorD}{HTML}{00FF00}

\begin{tikzpicture}

\begin{axis}[%
width=.8\columnwidth,
height=.6\columnwidth,
ymode=log,
xmode=log,
scale only axis,
xtick={2,8,32,128,512},
xlabel={Number of signatures},
xticklabels={$ 2 $, $ 8 $, $ 32 $, $ 128 $,$ 512$},
max space between ticks=20,
ylabel={Prover Time (seconds)},
ylabel near ticks,
yticklabel shift={0cm},
axis background/.style={fill=white},
legend columns=4,
legend style={legend cell align=left, align=left, fill=none, draw=none, font=\small,inner sep=-0pt, row sep=0pt},
legend pos = north west,
label style={font=\small},
tick label style={font=\small},
ytick style={draw=none},
ymajorgrids,
xmajorgrids,
grid style={line width=.5pt, draw=gray!90,dashed},
major grid style={line width=.2pt,draw=gray!50},
legend columns=1,
]

\addplot [color=colorC, mark=square*, mark options={scale=0.7,solid, colorC}]
  table[row sep=crcr, y expr=\thisrow{Y} * 2/8]{%
X Y
2 88.3013 \\
2 88.3013 \\
8 381.461616 \\
32 1638.872128 \\
128 7007.591168 \\
512 29838.7753 \\
};
\addplot [color=colorB, mark=square*, mark options={scale=0.7,solid, colorB}]
  table[row sep=crcr,y expr=\thisrow{Y} /8]{%
X Y
2 88.3013 \\
2 88.3013 \\
8 88.3013 \\
32 381.461616 \\
128 1638.872128 \\
512 7007.591168 \\
};
\addplot [color=colorA, mark=square*, mark options={scale=0.7,solid, colorA}]
  table[row sep=crcr,y expr=\thisrow{Y} /8]{%
X Y
2 88.3013 \\
2 88.3013 \\
8 88.3013 \\
32 88.3013 \\
128 381.461616 \\
512 1638.872128 \\
};

\addplot [color=colorD, mark=square*, mark options={scale=0.7,solid, colorD}]
  table[row sep=crcr,y expr=\thisrow{Y} /8]{%
X Y
2 88.3013 \\
2 88.3013 \\
8 88.3013 \\
32 88.3013 \\
128 88.3013 \\
512 381.461616 \\
};
\addlegendentry{The original Virgo}
\addlegendentry{$8$-machine \devirgo}
\addlegendentry{$32$-machine \devirgo}
\addlegendentry{$128$-machine \devirgo}
\end{axis}
\end{tikzpicture}%
    \end{subfigure}
    \caption{Prover time of \devirgo and the original Virgo for Cosmos block header verification.}
    \label{fig:virgo_competition}
\vspace{1cm}
\end{figure}
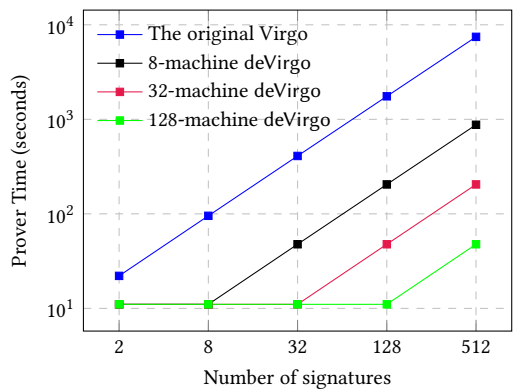

Figure~\ref{fig:virgo_competition} shows the prover time (in seconds) against different $N$. 
For $\devirgo$, we repeat the experiment with $8$, $32$, $128$ distributed machines.
According to~\cref{fig:virgo_competition}, the prover time of the original Virgo increases linearly in the number of signatures $N$, while the prover time of \devirgo is almost independent of $N$ until $N$ is greater than the number of servers when computation becomes an bottleneck. The linear scalability suggests that the workload of each machine only depends on its own sub-circuit and the communication overhead is small.  
\Cref{tab:eval} reports the communication cost among parallel machines.
The total communication cost is linear in the number of machines, consistent with the analysis in~\cref{subsec:distcombine}, with each machine sending and receiving around 1 GB of data.
Since we envision a relayer node in \systemname to be deployed in a data-center-like environment, the amount of traffic is reasonable.

In practice, the Cosmos block headers typically have $N=128$ signatures while $32$ top signatures are sufficient to achieve super-majority. Therefore, generating a correctness proof for a Cosmos block header would take more than 400 seconds with the original Virgo, but it decreases to 13.28 seconds with \devirgo, implying a 30x speedup.
In general, as is consistent with the analysis in~\cref{sec:dist}, \devirgo accelerates the proof generation on data-parallel circuits with $N$ copies by a factor of almost $N$, which is optimal for distributed algorithms.

\subsubsection{Proof size and verification time.}

To reduce on-chain verification cost, we use the recursive verification technique presented in~\cref{sec:composition}. Now we report on its efficacy. 

\paragraph{Recursive proof generation time.}
We implement recursive verification by invoking Groth16 (constructed using gnark~\cite{gnark}) on the verification circuit.
We report the proof time in \devirgo, the generation time of recursive proofs (the column marked RV), and the sum, in~\cref{tab:eval}, for various numbers of signatures.
The RV time almost remains constant in the number of signatures verified by the $\devirgo$ proofs.
That is because of the data-parallel structure of the state transition proof circuit: the size of Groth16 verification circuit is only a function of the size of a sub-circuit.

The main benefit of recursive verification is a reduction in both proof size and verification cost.

\paragraph{Reduced proof size.}
\Cref{tab:eval} shows the proof size both with and without recursive verification. 
For the practical scenario where $N=32$, the proof size is reduced from 1.9 MB to 131 Bytes. 
Overall, for $N=32$, with an increase of about $25\%$ in prover time, we get a reduction of around 14000x in proof size.

\paragraph{Reduced on-chain verification cost.}
The final proof is 131 Bytes while the final verification only costs $3$ pairings.
As shown in \cref{tab:eval}, the on-chain verification cost is constant (227K). 
In comparison, without recursive verification, directly verifying Virgo proofs on-chain would be infeasible. (Our estimation of the gas cost is 78M, which far exceeds the single block gas limit 30M). 

\subsubsection{Comparison with optimistic bridges.} With batching, the confirmation latency of \systemname is under $2$ minutes, including $3 \times 32$ seconds for waiting for all blocks in the batch and another $20$ seconds for proof generation. 
While this is not blazing fast, in comparison, optimistic bridges have much longer confirmation time. E.g., NEAR's Rainbow bridge has a challenge window of $4$ hours~\cite{nearbridge} before which the transfer cannot be confirmed.

\subsection{Cost analysis}
\label{sec:cost}
In this section, we analyze the operational cost of \systemname, which consists of off-chain cost (generating proofs) and on-chain cost (storing headers and verifying proofs).

\paragraph{Off-chain cost.}
Off-chain cost can vary significantly based on the deployment. While we use AWS in our performance benchmark, it may not be the best option for practical deployment. AWS service is expensive due to its high margin, elastic scaling capability, and high reliability, which isn't necessary for our proof generation process. To show a representative range, we consider two deployment options: cloud-based and self-hosted. For cloud-based deployment, we search for reputable and economical dedicated server rental services and choose Hetzner\cite{hetzner} as an example. For self-hosted options, we calculate the cost to purchase the hardware and the on-going cost (mainly the electricity).

On AWS c5.24xlarge, it takes $18$ seconds to generate a proof with $32$ machines.
Renting a server with a similar spec as AWS c5.24xlarge from Hetzner costs \$253.12 per month, thus the cost of cloud-based deployment with Hetzner will be around \$8100 per month for all 32 machines. It translates to \$0.02 per block.

To estimate the cost for self-hosted deployment, we use online tools to configure a machine with a comparable spec to that in AWS. \Cref{table:hwConfig} in~
\cref{app:machineconfig} reports the configuration and each machine costs around \$4.5k. The total setup cost is thus around \$4.5k $\times 32=\$144$k. For self-hosted servers, the main on-going cost is electricity. With each machine consuming $657$W power, a $32$-machine cluster consumes $0.105$ kWh per block. Assuming US average electricity rate $\$0.12$/KWh~\cite{USEnergy}, the electricity cost is \$$0.012$ per block, or \$5184 per month.

\paragraph{On-chain cost.} On-chain cost refers to the total gas used for on-chain operation, and we report the equivalent USD cost based on the gas price (about $20$ gwei) and ETH price (about $1600$ USD) at the time of writing (August 2022).
If we use efficient batched proofs, for a batch of $N$ headers, the bulk of the verification cost is that of verifying one Groth16 proof, which costs less than $230K$ gas, roughly \$$7.36$. If we choose $N=32$ for example, the on-chain cost will be \$$0.23$ per block. Moreover, if we adopt the optimization mentioned in appendix \ref{app:optimisc}, we can further reduce the on-chain cost and offload the cost to users if the number of users is large.

\subsection{Ethereum to other EVM-compatible chains}
\label{sec:theotherdir}

So far we have focused on the bridge from Cosmos to Ethereum because generating and verifying correctness proofs for that direction is challenging. We also implement a prototype of a bridge from Ethereum to other EVM-compatible blockchains.

The high level idea is simple: upon receiving a block header, the updater contract on the receiver chain verifies the PoW and appends it to the list of headers if the verification is passed.
However, a wrinkle to the implementation is that Ethereum uses a memory hard hash function, EthHash~\cite{wood2014ethereum}, which is prohibitively inefficient to run on-chain.
Basically, EthHash involves randomly accessing elements in a 1 gigabyte dataset (called a DAG) derived from a public seed and the block height. Generating the DAGs on-chain is prohibitively expensive. 

Our idea is to pre-compute many DAGs off-chain and store their hashes on-chain. Specifically, as part of \systemname setup, we pre-compute 2,048 DAGs , build a Merkle tree for each DAG using MiMC~\cite{albrecht2016mimc}, and store the Merkle roots on-chain. Per EthHash specification, a new DAG is generated every 30,000 blocks, so 2,048 of them can last for 10 years; the off-chain pre-computation process takes no more than 4 days.
Then, the correctness proofs will show that a given EthHash PoW is correct with respect to the Merkle root of the DAG corresponding to the block in question. We emphasize that the setup process is verifiable and anyone can verify the published Merkle roots on their own before using the service. The circuit for verifying EthHash PoW has around $2$ million gates. 

The rest of the protocol is the same as a regular light client, which involves storing the headers, following the longest chain by computing accumulated difficulty, resolving forks, etc.

\paragraph{Cost analysis.} Since EthHash PoW verification circuit has only around $2$ million constraints, a single machine with the configuration in Appendix \ref{app:machineconfig} can generate a proof within 10 seconds. 
As long as the receiver chain is EVM-compatible, the on-chain cost will be close to that presented in~\cref{sec:cost}, since the updater contract only verifies Groth16 proofs in all cases.

\section{Related work}

In this section, we compare \systemname to existing cross-chain bridge systems and the line of work on zk-rollups which also uses ZKPs for scalability and security.

\paragraph{Cross-chain bridges in the wild and security issues.}
Cross-chain systems are widely deployed and used. Below we briefly survey the representative ones. The list is not meant to be exhaustive. 
PolyNetwork~\cite{poly_network} is an interoperability protocol using a side-chain as the relay with a two-phase commitment protocol. Wormhole~\cite{wormhole} is a generic message-passing protocol secured by a network of guardian nodes, and its security relies on $\nicefrac{2}{3}$ of the committee being honest.
Ronin operates in a similar model.
While relying on decentralized committees for security, practical deployment usually opts for relatively small ones for efficiency (e.g., 9 in case of Ronin). Committee breaches are far from being rare in practice.
In a recent exploit against Ronin~\cite{RoninAttackCoindesk}, the attacker obtained five of the nine validator keys, stealing 624 million USD. PolyNetwork and Wormhole were also recently attacked, losing \$611m~\cite{polynetworkattack} and \$326m~\cite{wormholeattack} respectively. Key compromise was suspected in the PolyNetwork attack.

An alternative design is to leverage economic incentives. Nomad~\cite{nomad} (which recently lost more than \$190m to hackers due to an implementation bug~\cite{nomad_hack}) and Near's Rainbow Bridge~\cite{rainbow} are such examples. These systems require participants to deposit a collateral, and rely on a watchdog service to continuously monitor the blockchain and confiscate offenders' collateral upon detecting invalid updates. 
Optimistic protocols fundamentally require a long confirmation latency in order to ensure invalid updates can be detected with high probability (e.g., Near~\cite{rainbow} requires 4 hours).
Moreover, participants must deposit significantly collateral (e.g., 20 ETH in Near~\cite{rainbow}).
Both issues can be avoided by \systemname.

In summary, compared to existing protocols, \systemname achieve both efficiency and cryptographic assurance.
\systemname is ``trustless'' in that it does not require extra assumptions other than those of blockchains and underlying cryptographic protocols. 
It also avoids the long confirmation of optimistic protocols.

\paragraph{zk-rollups.}
Rollups are protocols that batch transaction execution using ZKPs to scale up the layer-1 blockchains. Starkware~\cite{starkware}, ZkSync~\cite{zksync}, and Polygon Zero~\cite{polygon_zero} are a few examples. 

These zk-rollup solutions have not been applied to the bridge setting, where our work is the first to use ZKP to enable a decentralized trustless bridge. In addition, the current zk-rollup work in general has not dealt with such large circuits as in \systemname, whereas in our work, we need to design and develop a number of techniques including \devirgo  and proof recursion to make building a ZKP-based bridge practical for the first time. In particular, we leverage the data parallelism of the circuits to obtain a ZKP protocol that is more than 100x faster than existing protocols for the workload in \systemname and combine it with proof recursion for efficient on-chain verification. The idea behind \devirgo protocol may be applicable to zk-rollups too.

\section*{Acknowledgments}
This material is in part based upon work supported by the National Science Foundation (NSF) under Grant No. TWC-1518899 and Grant No. 2144625, DARPA under Grant No. N66001-15-C-4066, the Center for Responsible, Decentralized Intelligence at Berkeley (Berkeley RDI), the Center for Long-Term Cybersecurity (CLTC), the Simons Foundation, and NTT Research. Any opinions, findings, and conclusions or recommendations expressed in this material are those of the author(s) and do not necessarily reflect the views of these institutes.
\bibliographystyle{ACM-Reference-Format}
\bibliography{ref.bib}
\appendix
\newpage

\section{Background: The sumcheck protocol}\label{sec:sumcheck}
The sumcheck protocol is given in Protocol~\ref{prot::sumcheck}.
\begin{figure}[t!]
\centering{\centering
\framebox{\parbox{.99\linewidth}{
\begin{protocol}[\textbf{Sumcheck}]
	\label{prot::sumcheck}
	The protocol proceeds in $\ell$ rounds. 
	\begin{itemize}
		\item In the first round, $\P$ sends a univariate polynomial $$f_1(x_1)\overset{def}{=}\sum\limits_{b_2,\ldots,b_\ell\in\{0,1\}}f(x_1,b_2,\ldots,b_\ell)\, ,$$ $\V$ checks $H=f_1(0)+f_1(1)$. Then $\V$ sends a random challenge $r_1\in\mathbb{F}$ to $\P$.
		\item In the $i$-th round, where $2\le i \le \ell-1$, $\P$ sends a univariate polynomial
		$$f_{i}(x_{i})\overset{def}{=}\sum\limits_{b_{i+1},\ldots,b_\ell\in\{0,1\}}f(r_1,\ldots, r_{i-1}, x_{i}, b_{i+1},\ldots, b_{\ell})\, ,$$ 
		$\V$ checks $f_{i-1}(r_{i-1})=f_{i}(0)+f_{i}(1)$, and sends a random challenge $r_{i}\in\mathbb{F}$ to $\P$.
		\item In the $\ell$-th round, $\P$ sends a univariate polynomial $$f_{\ell}(x_{\ell})\overset{def}{=}f(r_1, r_2, \ldots, r_{l-1}, x_{\ell})\, ,$$ $\V$ checks $f_{\ell-1}(r_{\ell-1})=f_{\ell}(0)+f_{\ell}(1)$. The verifier generates a random challenge $r_{\ell}\in\mathbb{F}$. Given oracle access to an evaluation $f(r_1, r_2, \ldots, r_\ell)$ of $f$, $\V$ will accept if and only if $f_{\ell}(r_\ell) = f(r_1, r_2, \ldots, r_\ell)$. The oracle access can be instantiated by $\PC$.
	\end{itemize}
\end{protocol}}}}
\end{figure}
\section{The distributed sumcheck protocol}\label{sec:distsumcheck}
The distributed sumcheck protocol is given in Protocol~\ref{prot::distsumcheck}.
\begin{figure*}[htbp]

\begin{mdframed}[nobreak = false]{
\begin{protocol}[\textbf{Distributed sumcheck}]
	\label{prot::distsumcheck}
    Suppose the prover has $N$ machines  $\P_0, \cdots, \P_{N-1}$ and suppose $\P_0$ is the master node. Each $P_i$ holds a polynomial $f^{(i)}: \F^{\ell-n} \rightarrow \F$ such that $f^{(i)}(\vecx) = f(\vecx[1:\ell-n], \veci)$. Suppose $\V$ is the verifier. The protocol proceeds in 3 phases consisting of $\ell$ rounds. 
	\begin{enumerate}
		\item\label{step: distsumcheck1} In the $j$-th round, where $1 \le j \le \ell-n$, each $\P_i$ sends $P_0$ a univariate polynomial
		$$f^{(i)}_j(x_j)\overset{def}{=}\sum\limits_{\vecb\in\{0,1\}^{\ell-n-j}}f^{(i)}(\vecr[1:j-1], x_j, \vecb)\, ,$$
		After receiving all univariate polynomials from $P_1, \cdots, P_{N-1}$, $P_0$ computes 
		$$f_j(x_j) = \sum\limits_{i = 0}^{N-1}f^{(i)}_j(x_{j})$$ then sends $f_j(x_j)$ to $\V$.  $\V$ checks $f_{j-1}(r_{j-1})=f_{j}(0)+f_{j}(1)$, and sends a random challenge $r_{j}\in\mathbb{F}$ to $\P_0$. $\P_0$ relays $r_j$ to $P_1, \cdots, P_{N-1}$.
		\item\label{step: distsumcheck2} In the $j$-th round, where $j= \ell-n$, after receiving $r_j$ from $\P_0$, each $\P_i$ computes $f^{(i)}(\vecr[1:j])$ and sends $f^{(i)}(\vecr[1:j])$ to $\P_0$. Then $\P_0$ constructs a multi-linear polynomial $f': \F^{n} \rightarrow \F$ such that $f'(\veci) = f^{(i)}(\vecr[1:j])$ for $0 \leq i < N$.
		\item\label{step: distsumcheck3} In the $j$-th round, where $\ell-n < j \le \ell$, $\P_0$ and $\V$ run Protocol~\ref{prot::sumcheck} on the statement:
		$$H' = \sum_{\vecb\in\{0, 1\}^n}f'(\vecb),$$
		where $H' = \sum_{i=0}^{N-1}f^{(i)}(\vecr[1:\ell-n])$.
	\end{enumerate}
\end{protocol}}
	\end{mdframed}
\end{figure*}
\section{Background: The GKR protocol}\label{sec:gkr}

\paragraph{Notations in GKR protocol. } We follow the convention in prior works of GKR protocols~\cite{CMT,t13,vsql,libra,virgo}. We denote the number of gates in the $i$-th layer as $S_i$ and let $s_i = \ceil{\log S_i}$. (For simplicity, we assume $S_i$ is a power of 2, and we can pad the layer with dummy gates otherwise.) We then define a function $V_i:\binary^{s_i}\rightarrow\mathbb{F}$ that takes a binary string $\vecb\in\binary^{s_i}$ and returns the output of gate $\vecb$ in layer $i$, where $\vecb$ is called the gate label. With this definition, $V_0$ corresponds to the output of the circuit, and $V_d$ corresponds to the input layer. Finally, we define two additional functions $add_i, mult_i: \binary^{s_{i-1}+2s_i}\rightarrow\binary$, referred to as \emph{wiring predicates} in the literature. $add_i$ ($mult_i$) takes one gate label $\vecz\in\binary^{s_{i-1}}$ in layer $i-1$ and two gate labels $\vecx,\vecy\in\binary^{s_i}$ in layer $i$, and outputs 1 if and only if gate $\vecz$ is an addition (multiplication) gate that takes the output of gate $\vecx,\vecy$ as input. With these definitions, for any $\vecz\in\binary^{s_i}$, $V_i$ can be written as:

\begin{equation}\label{eq:naiveGKR}
    \begin{aligned}
	V_i(\vecz)=\sum\nolimits_{\vecx, \vecy \in\binary^{s_{i+1}}}(add_{i+1}(\vecz,\vecx,\vecy)(V_{i+1}(\vecx)+V_{i+1}(\vecy))\\
	+mult_{i+1}(\vecz,\vecx,\vecy)V_{i+1}(\vecx)V_{i+1}(\vecy))
	\end{aligned}
\end{equation}

In the equation above, $V_i$ is expressed as a summation, so $\V$ can use the sumcheck protocol to check that it is computed correctly. As the sumcheck protocol operates on polynomials defined on $\mathbb{F}$, we rewrite the equation with their multi-linear extensions:
\begin{align}\label{eq:appGKR}
\tV_i(\vecg)=&\sum\nolimits_{\vecx, \vecy \in\binary^{s_{i+1}}}h_i(\vecg, \vecx, \vecy)\nonumber\\
=&\sum\nolimits_{\vecx, \vecy \in\binary^{s_{i+1}}}(\tadd_{i+1}(\vecg,\vecx,\vecy)(\tV_{i+1}(\vecx)+\tV_{i+1}(\vecy))\nonumber\\
&+\tmult_{i+1}(\vecg,\vecx,\vecy)\tV_{i+1}(\vecx)\tV_{i+1}(\vecy))\,,
\end{align}
where $\vecg\in\mathbb{F}^{s_i}$ is a random vector. 

\paragraph{The GKR Protocol.}With Equation~\ref{eq:appGKR}, the GKR protocol proceeds as following. The prover $\P$ first sends the claimed output of the circuit to $\V$. From the claimed output, $\V$ defines polynomial $\tV_0$ and computes $\tV_0(\vecg)$ for a random $\vecg\in\mathbb{F}^{s_0}$. $\V$ and $\P$ then invoke a sumcheck protocol on Equation~\ref{eq:appGKR} with $i=0$. As described in Protocol~\ref{prot::sumcheck}, at the end of the sumcheck, $\V$ needs an oracle access to $h_i(\vecg,\vecu,\vecv)$, where $\vecu,\vecv$ are randomly selected in $\mathbb{F}^{s_{i+1}}$. To compute $h_i(\vecg,\vecu,\vecv)$, $\V$ computes $\tadd_{i+1}(\vecg,\vecu,\vecv)$ and $\tmult_{i+1}(\vecg,\vecu,\vecv)$ locally (they only depend on the wiring pattern of the circuit, not on the values), asks $\P$ to send $\tV_1(\vecu)$ and $\tV_1(\vecv)$ and computes $h_i(\vecg,\vecu,\vecv)$ to complete the sumcheck protocol. In this way, $\V$ and $\P$ reduce a claim about the output to two claims about values in layer 1. $\V$ and $\P$ could invoke two sumcheck protocols on $\tV_1(\vecu)$ and $\tV_1(\vecv)$ recursively to layers above, but the number of the sumcheck protocols would increase exponentially.

\smallskip\noindent\textbf{Combining two claims using a random linear combination.}  One way to combine two claims $\tV_i(\vecu)$ and $\tV_i(\vecv)$ is using random linear combinations, as proposed in~\cite{zksumcheck,hyrax}. Upon receiving the two claims $\tV_i(\vecu)$ and $\tV_i(\vecv)$, $\V$ selects $\alpha_{i,1}, \alpha_{i,2}\in\mathbb{F}$ randomly and computes $\alpha_{i, 1}\tV_i(\vecu)+\alpha_{i, 2}\tV_i(\vecv)$. Based on Equation~\ref{eq:appGKR}, this random linear combination can be written as
{%
\begin{align}\label{eq:randomGKR}
&\alpha_{i, 1}\tV_i(\vecu)+\alpha_{i, 2}\tV_i(\vecv)\nonumber\\
=&\alpha_{i, 1}\sum_{\vecx, \vecy \in\binary^{s_{i+1}}}\tadd_{i+1}(\vecu,\vecx,\vecy)(\tV_{i+1}(\vecx)+\tV_{i+1}(\vecy))\nonumber\\+&\tmult_{i+1}(\vecu,\vecx,\vecy)\tV_{i+1}(\vecx)\tV_{i+1}(\vecy)\nonumber\\+&\alpha_{i, 2}\sum_{\vecx, \vecy \in\binary^{s_{i+1}}}\tadd_{i+1}(\vecv,\vecx,\vecy)(\tV_{i+1}(\vecx)+\tV_{i+1}(\vecy))\nonumber\\+&\tmult_{i+1}(\vecv,\vecx,\vecy)\tV_{i+1}(\vecx)\tV_{i+1}(\vecy)\nonumber\\
=&\sum_{\vecx, \vecy \in\binary^{s_{i+1}}}(\alpha_{i, 1}\tadd_{i+1}(\vecu,\vecx,\vecy)+\alpha_{i, 2}\tadd_{i+1}(\vecv,\vecx,\vecy))(\tV_{i+1}(\vecx)+\tV_{i+1}(\vecy))\nonumber\\
&+(\alpha_{i, 1}\tmult_{i+1}(\vecu,\vecx,\vecy)+\alpha_{i, 2}\tmult_{i+1}(\vecv,\vecx,\vecy))\tV_{i+1}(\vecx)\tV_{i+1}(\vecy)
\end{align}}$\V$ and $\P$ then execute the sumcheck protocol on Equation~\ref{eq:randomGKR} instead of Equation~\ref{eq:appGKR}. At the end of the sumcheck protocol, $\V$ still receives two claims about $\tV_{i+1}$, computes their random linear combination and proceeds to the layer above recursively until the input layer.

The formal GKR protocol is presented in Protocol~\ref{prot:GKR}.

\begin{figure*}[t!]
	\begin{mdframed}[nobreak = false]
		
		\begin{protocol}[\textbf{GKR}]
			\label{prot:GKR}
			Let $\mathbb{F}$ be a finite field. Let $C$: $\mathbb{F}^m\rightarrow \mathbb{F}^k$ be a $d$-depth layered arithmetic circuit. $\mathcal{P}$ wants to convince that $C(\vecx) = \vecone$ where $\vecx$ is the input from $\V$, and $\vecone$ is the output. Without loss of generality, assume $m$ and $k$ are both powers of 2 and we can pad them if not.
			
			\begin{enumerate}
				\item\label{step:gkr1} $\V$ chooses a random $\vecg \in \mathbb{F}^{s_0}$ and sends it to $\P$.
				\item\label{step:gkr2} $\P$ and $\V$ run a sumcheck protocol on
				\[
				1=\sum_{\vecx, \vecy\in \binary^{s_{1}}}(\tadd_{1}(\vecg^{(0)}, \vecx, \vecy)(\tV_{1}(\vecx)+\tV_{1}(\vecy))+\tmult_{1}(\vecg^{(0)},\vecx,\vecy)\tV_{1}(\vecx)\tV_{1}(\vecy))
				\]
				At the end of the protocol, $\V$ receives $\tV_1(\vecu^{(1)})$ and $\tV_1(\vecv^{(1)})$. $\V$ computes $\tmult_1(\vecg^{(0)},\vecu^{(1)},\vecv^{(1)})$, $\tadd_1(\vecg^{(0)},\vecu^{(1)},\vecv^{(1)})$ and checks that $\tadd_1(\vecg^{(0)},\vecu^{(1)},\vecv^{(1)})$ $(\tV_1(\vecu^{(1)})+\tV_1(\vecv^{(1)}))+\tmult_1(\vecg^{(0)},\vecu^{(1)},\vecv^{(1)})$ $\tV_1(\vecu^{(1)})\tV_1(\vecv^{(1)})$ equals to the last message of the sumcheck.
				
				\item\label{step:gkr3} For $i=1,...,d-1$:
				\begin{itemize}
					
					\item $\V$ randomly selects $\alpha_{i,1}, \alpha_{i,2}\in\F$ and sends them to $\mathcal{P}$.
					\item $\P$ and $\V$ run the sumcheck on the equation\\
				
					\begin{align*}
						\alpha_{i, 1}\tV_i(\vecu^{(i)})+\alpha_{i, 2}\tV_i(\vecv^{(i)})=\sum_{\vecx, \vecy\in \binary^{s_{i+1}}}&((\alpha_{i, 1}\tilde{add}_{i+1}\vecu^{(i)}, \vecx, \vecy)+\alpha_{i, 2}\tilde{add}_{i+1}(\vecv^{(i)}, \vecx, \vecy))(\tV_{i+1}(\vecx)+\tV_{i+1}(\vecy))\\
						+&(\alpha_{i, 1}\tilde{mult}_{i+1}(\vecu^{(i)}, \vecx, \vecy)+\alpha_{i, 2}\tilde{mult}_{i+1}(\vecv^{(i)}, \vecx, \vecy))\tV_{i+1}(\vecx)\tV_{i+1}(\vecy))\nonumber
					\end{align*}
					
					\item At the end of the sumcheck protocol, $\P$ sends $\V$ $\tV_{i+1}(\vecu^{(i+1)})$ and $\tV_{i+1}(\vecv^{(i+1)})$.

					\item $\V$ computes the following and checks if it equals to the last message of the sumcheck. For simplicity, let $Mult_{i+1}(\vecx) = \tilde{mult}_{i+1}(\vecx, \vecu^{(i+1)}, \vecv^{(i+1)})$ and $Add_{i+1}(\vecx) = \tilde{add}_{i+1}(\vecx, \vecu^{(i+1)}, \vecv^{(i+1)})$.
					\begin{align*}
						&(\alpha_{i, 1}Mult_{i+1}(\vecu^{(i)})+\alpha_{i, 2}Mult_{i+1}(\vecv^{(i)})(\tV_{i+1}(\vecu^{(i+1)})\tV_{i+1}(\vecv^{(i+1)}))+\\
						&(\alpha_{i, 1}Add_{i+1}(\vecu^{(i)})+\alpha_{i, 2}Add_{i+1}(\vecv^{(i)})(\tV_{i+1}(\vecu^{(i+1)})+\tV_{i+1}(\vecv^{(i+1)}))
					\end{align*}
					If all checks in the sumcheck pass, $V$ uses $\tV_{i+1}(\vecu^{(i+1)})$ and $\tV_{i+1}(\vecv^{(i+1)})$ to proceed to the $(i+1)$-th layer. Otherwise, $\V$ outputs $\reject$ and aborts.
					
				\end{itemize}
				\item\label{step:gkr4} At the input layer $d$, $\V$ has two claims $\tV_{d}(\vecu^{(d)})$ and $\tV_{d}(\vecv^{(d)})$. $\V$ evaluates $\tV_d$ at $\vecu^{(d)}$ and $\vecv^{(d)}$ using the input and checks that they are the same as the two claims. If yes, output $\accept$;  otherwise, output $\reject$.
				
			\end{enumerate}
		\end{protocol}
	\end{mdframed}
\end{figure*}
\begin{table*}[!]
\begin{tabular}{|l|l|l|l|l|}
\hline
       Hardware type & Hardware name & Power consumption              & Price & Quantity \\
\hline
CPU & AMD Ryzen Threadripper 3970X & 435W & \$2325.99 & 1\\
\hline 
Memory & CMK256GX4M8D3600C18 & 96W & \$1129.99 & 1\\
\hline 
Motherboard & MSI TRX40 PRO WIFI & 80W & \$565.57 & 1\\
\hline
Power Supply & EVGA 220-T2-1000-X1 & 94\% efficiency & \$332.88 & 1 \\
\hline 
SSD & MZ-V8P1T0B/AM & 6.2W & \$129.99 & 1\\
\hline
Total & & 657W & \$4484.42 & \\
\hline
\end{tabular}
\vspace{.5cm}
\caption{Prover hardware configuration.\label{table:hwConfig}}
\end{table*}

\section{The distributed PC protocol}\label{sec:distpcprotocolonly}
The formal protocol of distributed polynomial commitment is given in Protocol~\ref{prot::distpc}.
\begin{figure*}[t!]
\centering{\centering
\framebox{\parbox{.99\linewidth}{
\begin{protocol}[\textbf{Distributed PC}]
	\label{prot::distpc}
    Suppose the prover has $N$ machines of $\P_0, \cdots, \P_{N-1}$ and suppose $\P_0$ is the master node. Each $P_i$ holds a polynomial $f^{(i)}: \F^{\ell-n} \rightarrow \F$ such that $f(\vecx) = \tbeta(\vecx[\ell-n+1:\ell], \veci)f^{(i)}(x[1:\ell-n])$. Suppose $\V$ is the verifier. Let $\H$ and $\L$ be two disjoint multiplicative subgroups of $\F$ such that $|\H| = \frac{2^\ell}{N}$ and $|\L| = \rho|\H|$. For simplicity, We assume $\rho = 1$. Let $\pp = \PC.\KeyGen(1^\lambda)$. The protocol proceeds in following steps.
	\begin{enumerate}
	    \item\label{step:distpc1} Each $\P_i$ invokes $\PC.\Commit(f^{(i)}, \pp)$ to compute $\vecf^{(i)}_\L$ by IFFT and FFT.  
		\item\label{step:distpc2} Each $\P_i$ sends $\vecf^{(i)}_\L[1]$, $\cdots$, $\vecf^{(i)}_\L[N]$ to $\P_0$, $\cdots$, $\P_{N-1}$ separately.
		\item\label{step:distpc3} Each $\P_i$ receives $\vecf^{(0)}_\L[i+1]$, $\cdots$, $\vecf^{(N-1)}_\L[i+1]$ from other machines. Assuming $\vech^{(i)}_\L$ = ($\vecf^{(0)}_\L[i+1]$, $\cdots$, $\vecf^{(N-1)}_\L[i+1])$, $\P_i$ computes $\com_{h^{(i)}}$ = $\MT.\Commit$($\vech^{(i)}_\L)$ and sends $\com_{h^{(i)}}$ to $\P_0$.
		\item\label{step:distpc4} Suppose $\vech$ = ($\com_{h^{(0)}}$, $\cdots$, $\com_{h^{(N-1)}})$, $\P_0$ computes $\com$ = $\MT.\Commit$($\vech)$ and sends $\com$ to $\V$.
		\item\label{step:distpc5} After receiving the random vector $\vecr$ from $\V$, $\P_0$ relays $\vecr$ to each $\P_i$. Each $\P_i$ computes $f^{(i)}(\vecr[1:\ell - n])$ and sends it to $\V$ via $\P_0$.
		\item\label{step:distpc6} To prove the correctness of $f^{(i)}(\vecr[1:\ell - n])$, given random index of $k_1, \cdots, k_c$ from $\V$, %
		$\P_{k_1-1}$, $\cdots$, $\P_{k_c-1}$ send $\vech^{(k_1-1)}_\L$, $\cdots$, $\vech^{(k_c-1)}_\L$ to $\V$ via $\P_0$.$\P_0$ also generates $(\vech[k_1], \pi_{k_1})$ = $\MT.\Open(\vech, k_1)$, $\cdots$, $(\vech[k_c], \pi_{k_c})$ = $\MT.\Open(\vech, k_c)$ and send them to $\V$.
        \item\label{step:distpc7} $\V$ checks $f(\vecr) = \sum_{i = 0}^{N-1}\tbeta(\vecr[\ell -n  + 1:\ell], \veci )f^{(i)}(\vecr[1:\ell - n])$.
        $\V$ checks $\vech[k_1]$=$\MT.\Commit(\vech^{(k_1-1)}_\L)$, $\cdots$, $\vech[k_c]$=$\MT.\Commit(\vech^{(k_c-1)}_\L)$. 
        Then $\V$ checks $\pi_{k_1}, \cdots, \pi_{k_c}$ by $\MT.\Verify(\pi_{k_1}, \vech[k_1],\com)$, $\cdots$, $\MT.\Verify(\pi_{k_c}, \vech[k_c],,\com)$. $\V$ also checks
        $q(\vecf^{(i)}_\L[k_1], \cdots, \vecf^{(i)}_\L[k_c], \vecf^{(i)}(\vecr[1:\ell-n])) = 0$ for each $i$ as shown in $\PC.\Verify$. If all checks pass, $\V$ outputs $\accept$, otherwise $\V$ outputs $\reject$.
		
	\end{enumerate}
\end{protocol}}}}
\end{figure*}

\section{Background: The Virgo protocol}\label{sec:virgo}
By combining the GKR protocol and the polynomial commitment in Section~\ref{subsec:distpc} We present the formal protocol of Virgo in Protocol~\ref{prot:virgo} and the the complexity of Protocol~\ref{prot:virgo} in the following\footnote{Protocol~\ref{prot:virgo} is a knowledge argument system rather than a zero-knowledge proof protocol as we actually use the knowledge argument system in our construction. }.

\paragraph{Complexity of Virgo~\cite{virgo}. }Given a layered arithmetic circuit $C$ with $d$ layers and $m$ inputs, Protocol~\ref{prot:virgo} is a zero-knowledge proof protocol as defined in Definition~\ref{def::zkp} for the function computed by $C$. The prover time is $O(|C| + m\log m)$. The proof size is $O(d\log |C|+\lambda\log^2 m)$ and The verification time is also $O(d\log |C|+\lambda\log^2 m)$.

\begin{figure*}[t!]
	\begin{mdframed}[nobreak = false]
		
		\begin{protocol}[\textbf{Virgo}]
			\label{prot:virgo}
			Let $\mathbb{F}$ be a finite field. Let $C$: $\mathbb{F}^m\rightarrow \mathbb{F}^k$ be a $d$-depth layered arithmetic circuit. $\mathcal{P}$ wants to convince that $\vecone=C(\vecx, \vecw)$ where $\vecx$ and $\vecw$ are input and $\vecone$ is the output. Without loss of generality, assume $m$ and $k$ are both powers of 2 and we can pad them if not.
			
			\begin{enumerate}
			    \item Set $\pp \leftarrow \PC.\KeyGen(1^\lambda)$. $\P$ invokes $\PC.\Commit(\tV_d, \pp)$ to generate $\com_{\tV_d}$ and sends $\com_{\tV_d}$ to $\V$.
			    \item $\P$ and $\V$ run step~\ref{step:gkr1}-\ref{step:gkr3} in Protocol~\ref{prot:GKR}. 
				\item At the input layer $d$, $\V$ has two claims $\tV_{d}(\vecu^{(d)})$ and $\tV_{d}(\vecv^{(d)})$. $\P$ and $\V$ invoke $\PC.\Open$ and $\PC.\Verify$ on $\tV_{d}(\vecu^{(d)})$ and  $\tV_{d}(\vecv^{(d)})$ with $\com_{\tV_d}$ and $\pp$. If they are equal to $\tV_{d}(\vecu^{(d)})$ and  $\tV_{d}(\vecv^{(d)})$ sent by $\P$, $\V$ outputs $\accept$, otherwise $\V$ outputs $\reject$.
				
			\end{enumerate}
		\end{protocol}
	\end{mdframed}
\end{figure*}

\section{The distributed Virgo protocol}\label{sec:distvirgo}
By combining Protocol~\ref{prot::distpc} and Protocol~\ref{prot::distsumcheck}, we present the formal protocol of \devirgo in Protocol~\ref{prot:distvirgo}. 
\begin{figure*}[t!]
	\begin{mdframed}[nobreak = false]
		
		\begin{protocol}[\textbf{Distributed Virgo}]
			\label{prot:distvirgo}
			Let $\mathbb{F}$ be a finite field. Let $C$: $\mathbb{F}^{mN}\rightarrow \mathbb{F}^k$ be a $d$-depth layered arithmetic circuit. Suppose $C$ is also a data-parallel circuit with $N$ identical copies. $\mathcal{P}$ is a prover with $N$ distributed machines and wants to convince $\V$ that $\vecone=C(\vecx, \vecw)$ where $\vecx$ and $\vecw$ are input, and $\vecone$ is the output. Without loss of generality, assume $m$, $N$, and $k$ are powers of 2 and we can pad them if not.
			
			\begin{enumerate}
			    \item Set $\pp \leftarrow \PC.\KeyGen(1^\lambda)$. Define the multi-linear extension of array $(\vecx, \vecw)$ as $\tV_d$. $\P$ invokes step~\ref{step:distpc1}-\ref{step:distpc4} in Protocol~\ref{prot::distpc} on $\tV_d$ to get $\com_{\tV_d}$ and sends $\com_{\tV_d}$ to $\V$.
				\item Define the multi-linear extension of array $\vecone$ as $\tV_0$. $\V$ chooses a random $g \in \mathbb{F}^{s_0}$ and sends it to $\P$. %
				\item $\P$ and $\V$ run Protocol~\ref{prot::distsumcheck}, the distributed sumcheck protocol, on
				\[
				1=\sum_{\vecx, \vecy\in \binary^{s_{1}}}(\tadd_{1}(\vecg^{(0)}, \vecx, \vecy)(\tV_{1}(\vecx)+\tV_{1}(\vecy))+\tmult_{1}(\vecg^{(0)},\vecx,\vecy)\tV_{1}(\vecx)\tV_{1}(\vecy))
				\]
				At the end of the protocol, $\V$ receives $\tV_1(\vecu^{(1)})$ and $\tV_1(\vecv^{(1)})$. $\V$ computes $\tmult_1(\vecg^{(0)},\vecu^{(1)},\vecv^{(1)})$, $\tadd_1(\vecg^{(0)},\vecu^{(1)},\vecv^{(1)})$ and checks that $\tadd_1(\vecg^{(0)},\vecu^{(1)},\vecv^{(1)})$ $(\tV_1(\vecu^{(1)})+\tV_1(\vecv^{(1)}))+\tmult_1(\vecg^{(0)},\vecu^{(1)},\vecv^{(1)})$ $\tV_1(\vecu^{(1)})\tV_1(\vecv^{(1)})$ equals to the last message of the sumcheck.
				
				\item For $i=1,...,d-1$:
				\begin{itemize}
					
					\item $\V$ randomly selects $\alpha_{i,1}, \alpha_{i,2}\in\F$ and sends them to $\mathcal{P}$.
					\item $\P$ and $\V$ run Protocol~\ref{prot::distsumcheck}, the distributed sumcheck protocol, on\\

					\begin{align*}
						\alpha_{i, 1}\tV_i(\vecu^{(i)})+\alpha_{i, 2}\tV_i(\vecv^{(i)})=\sum_{\vecx, \vecy\in \binary^{s_{i+1}}}&((\alpha_{i, 1}\tilde{add}_{i+1}(\vecu^{(i)}, \vecx, \vecy)+\alpha_{i, 2}\tilde{add}_{i+1}(\vecv^{(i)}, \vecx, \vecy))(\tV_{i+1}(\vecx)+\tV_{i+1}(\vecy))\\
						+&(\alpha_{i, 1}\tilde{mult}_{i+1}(\vecu^{(i)}, \vecx, \vecy)+\alpha_{i, 2}\tilde{mult}_{i+1}(\vecv^{(i)}, \vecx, \vecy))\tV_{i+1}(\vecx)\tV_{i+1}(\vecy))\nonumber
					\end{align*}
					
					\item At the end of the distributed sumcheck protocol, $\P$ sends $\V$ $\tV_{i+1}(\vecu^{(i+1)})$ and $\tV_{i+1}(\vecv^{(i+1)})$.

					\item $\V$ computes the following and checks if it equals to the last message of the sumcheck. For simplicity, let $Mult_{i+1}(\vecx) = \tilde{mult}_{i+1}(\vecx, \vecu^{(i+1)}, \vecv^{(i+1)})$ and $Add_{i+1}(\vecx) = \tilde{add}_{i+1}(\vecx, \vecu^{(i+1)}, \vecv^{(i+1)})$.
					\begin{align*}
						&(\alpha_{i, 1}Mult_{i+1}(\vecu^{(i)})+\alpha_{i, 2}Mult_{i+1}(\vecv^{(i)})(\tV_{i+1}(\vecu^{(i+1)})\tV_{i+1}(\vecv^{(i+1)}))+\\
						&(\alpha_{i, 1}Add_{i+1}(\vecu^{(i)})+\alpha_{i, 2}Add_{i+1}(\vecv^{(i)})(\tV_{i+1}(\vecu^{(i+1)})+\tV_{i+1}(\vecv^{(i+1)}))
					\end{align*}
					If all checks in the sumcheck pass, $V$ uses $\tV_{i+1}(\vecu^{(i+1)})$ and $\tV_{i+1}(\vecv^{(i+1)})$ to proceed to the $(i+1)$-th layer. Otherwise, $\V$ outputs $\reject$ and aborts.
					
				\end{itemize}
				\item At the input layer $d$, $\V$ has two claims $\tV_{d}(\vecu^{(d)})$ and $\tV_{d}(\vecv^{(d)})$. $\P$ invokes step~\ref{step:distpc5}-\ref{step:distpc6} in Protocol~\ref{prot::distpc} to open $\tV_{d}(\vecu^{(d)})$ and  $\tV_{d}(\vecv^{(d)})$ while $\V$ invokes step~\ref{step:distpc7} in Protocol~\ref{prot::distpc} to validate $\tV_{d}(\vecu^{(d)})$ and  $\tV_{d}(\vecv^{(d)})$. If they are equal to $\tV_{d}(\vecu^{(d)})$ and  $\tV_{d}(\vecv^{(d)})$ sent by $\P$, $\V$ outputs $\accept$, otherwise $\V$ outputs $\reject$.
				
			\end{enumerate}
		\end{protocol}
	\end{mdframed}
\end{figure*}

\section{On-chain Gas Cost Optimization}  \label{app:optimisc}
To further optimize the on-chain gas cost of block header verification and storage for a universal zkBridge, we propose the following approach, in which the prover won't bother to pay for on-chain proof verification or block header storage, and users are encouraged to submit the proof they need by our incentive design.

In our optimization, the same as the aforementioned batched proof, the prover generates one single proof for every $2^d$ blocks where $d$ is a system configuration, and each proof checks and shows the validity of all signatures in the corresponding $2^d$ blocks. However, instead of submitting the Merkle root of the batch along with the proof on-chain immediately, provers simply post the proof to the users (e.g., through a website), and it's up to the users to retrieve and post the proof on-chain. Thus there's no more on-chain gas cost for provers through the approach.

For users who want to verify a transaction $tx$ in a block $blk$, the workflow is as follows.

\begin{enumerate}
    \item If $blk$ has already been submitted on-chain, go to the next step. Otherwise, retrieve the proofs for the sequence of blocks from the first unsubmitted one to $blk$, and then invoke the updater contract to verify all the proofs on-chain and store the information of the corresponding sequence of blocks. The process can be expensive. However, once the proofs are verified and the blocks are confirmed by the updater contract, the user becomes the owner of all these proofs on-chain, and can benefit from the proofs by charging later users who rely on these proofs to verify their transactions on-chain.
    
    \item  Thanks to the previously submitted proofs, the validity of the corresponding block is already proved at this step. And the work can never be accomplished without the efforts of proving all the blocks prior to $blk$ (including $blk$). Suppose $blk$ is the $i^{th}$ block, then for each block with index in the range $[i-t+1, i]$, the user should pay a certain amount of fee to the block proof owner in compensation, where $t$ is a system configuration and the definition of block proof owner is defined in the previous step.
    
\end{enumerate}

In this case, provers don't bother to pay for on-chain verification any more, and the proofs are only submitted and verified on demand, which is more cost-efficient and can reduce possible waste. Moreover, through carefully-designed incentive, we can actually encourage users to submit the proofs as a possible investment, and it can also help with the popularity of our bridge.

Through the optimization, the cost performance of our bridge can be summarized as follows. If there is high demand, then each proof will be submitted immediately upon generation, and in this case each user needs to pay for at most one time of on-chain proof verification. It then degenerates into our original batched proof, but users are responsible of paying for the on-chain verification instead. If the sender chain is so unpopular that there is little bridging demand from the chain, then we successfully avoid unnecessarily submitting the proofs on-chain for meaningless but costly verification. And even if a user suddenly exists and requires bridging in this case, the request can also be fulfilled by retrieving the proofs from provers and sending them for on-chain verification one by one.

And thus we can see that, the new design can actually benefit both the provers and the users.

\section{Prover machine configuration} \label{app:machineconfig}
To estimate the power consumption, we simulated a computer build. The total power consumption are based on the spec provided by the manufacturer. We present our hardware configuration in Table \ref{table:hwConfig}. Our prover machine doesn't need to be highly reliable since proof generation can be interrupted and restart at any time so we choose consumer grade hardware to be cost effective.

\end{document}